\documentclass[11pt]{article}

\usepackage[margin=1in]{geometry}
\usepackage{amsmath}
\usepackage{amssymb}
\usepackage{graphicx}
\usepackage{subcaption}
\usepackage{dcolumn}
\usepackage{bm}
\usepackage{mathptmx}
\usepackage{etoolbox}
\usepackage{multirow}
\usepackage{tabularx}
\usepackage{makecell}
\usepackage{booktabs}
\usepackage{colortbl}
\usepackage{siunitx}
\usepackage{comment}
\usepackage{xcolor}

\usepackage{hyperref}
\usepackage{cleveref}

\graphicspath{{figures/}}
\makeatletter
\newcommand{\articletype}[1]{}
\newcommand{\orcid}[1]{}
\newcommand{\affil}[1]{\g@addto@macro\@author{\\[2pt]\small #1}}
\newcommand{\email}[1]{\g@addto@macro\@author{\\[2pt]\small\texttt{#1}}}
\newcommand{\keywords}[1]{\gdef\@keywords{#1}}
\newcommand{\printkeywords}{%
  \ifdefined\@keywords\par\noindent\small\textbf{Keywords: }\@keywords\par\fi}
\makeatother
\newcommand{\affilskip}{\affil{\vspace{0pt}}}

\usepackage{fancyhdr}
\fancypagestyle{firstpage}{\fancyhf{}%
  \fancyfoot[C]{\footnotesize This is the version of the article before peer
  review or editing, as submitted by an author to Machine Learning: Engineering.
  IOP Publishing Ltd is not responsible for any errors or omissions in this
  version of the manuscript or any version derived from it.}}

\begin{document}

\title{Transferable Low-Dimensional Representations of Aircraft Surface Fields}

\author{Francis G. VanGessel$^{1,*,\dagger}$\orcid{0000-0003-4765-6499}, Cashen Diniz$^2$\orcid{https://orcid.org/0000-0001-6734-7111} and Mark Fuge$^{2}$\orcid{https://orcid.org/0000-0003-3819-8895}}

\affil{$^1$Department of Mechanical Engineering, University of Maryland, College Park, Maryland 20742, USA}

\affil{$^2$Department of Mechanical and Process Engineering, ETH Z\"urich, 8092 Z\"urich, Switzerland
}
\affilskip
\affil{$^*$Author to whom any correspondence should be addressed. E-mail: f.vangessel@colostate.edu}
\affil{$^{\dagger}$Present address: Department of Mechanical Engineering, Colorado State University, }
\affil{Fort Collins, Colorado 80523, USA}

\keywords{representation learning, transfer learning, aerospace analysis, aircraft surface fields}
\date{}
\maketitle

\thispagestyle{firstpage}

\begin{abstract}
Analyzing flow fields on aerodynamic surfaces is critical to designing next-generation aircraft. Raw computational representations are high-dimensional and costly for complex three-dimensional bodies, while conventional low-dimensional representations apply only to the systems from which they derive. We introduce an approach that learns fluid flow representations from simple two-dimensional geometries and transfers them to complex real-world three-dimensional aircraft. Comparing the recently proposed Least Volume autoencoder (LVAE) to proper orthogonal decomposition (POD), we show both yield compact 28–41 dimensional representations of two-dimensional flow fields, with the nonlinear LVAE producing more accurate in-domain reconstructions at matched dimensionality. These models transfer zero-shot to three-dimensional systems, and counterintuitively the linear POD transfers more accurately than LVAE. Applied to extruded wings, POD-reconstructed lift and drag remain within 0.1\% and 0.7\% of true values, while transfer to complex blended wing body aircraft preserves lift and drag to within 0.3\% and 0.7\% without any three-dimensional training data. We further show that three-dimensional flow patterns missed by the pretrained model can be isolated to train a disentangled 3D latent space via a dual encoder-decoder transfer learning strategy, performing accurately in previously inaccessible data-scarce regimes. Using only five blended wing training cases, mean squared reconstruction error drops by 5$\times$ relative to the 2D pretrained models and 70$\times$ relative to models trained from scratch. These results establish that transfer learning from readily available two-dimensional datasets unlocks compact, interpretable representations of complex three-dimensional aircraft surface flows, enabling design insights in data-starved early phases where conventional approaches require hundreds to thousands of simulations.\\
\printkeywords
\end{abstract}

\section{\label{sec:introduction}Introduction}

Modeling the surface flow over complex bodies allows scientists and engineers to interpret the aerodynamic behavior of the next generation of passenger aircraft, rotor vehicles, UAVs, and wind turbine designs. Understanding of the aerodynamic characteristics of these systems guides the design process by providing actionable insights into the rich interplay between geometry and aerodynamic performance. However, modern aircraft design faces two fundamental challenges. First, the high-dimensional flow representation inherent to the meshes, grids, or voxels used by high-fidelity computational fluid dynamics (CFD) simulations obscures salient flow patterns. Without compact, low-dimensional representations, direct interpretation of the relationship between geometric design choices and aerodynamic behavior becomes prohibitively difficult. Second, the data-scarce regimes inherent to early-stage design efforts provide insufficient information regarding the interplay between the high-dimensional geometric design space and aerodynamic performance. The fundamental limitation is therefore the tension between the low-dimensional representations required for actionable design insight, which require large datasets to reliably learn, and the data scarcity of new design efforts where such datasets do not yet exist. To overcome this challenge, we must demonstrate that interpretable low-dimensional representations of surface flow fields derived from simple, readily obtainable aerodynamic datasets can be effectively transferred to complex real-world aircraft design efforts with limited data while preserving critical performance metrics. Such transferable representations address both interpretability and data challenges of high-fidelity aircraft analysis.

Researchers have applied representation learning and the closely related approach of modal analysis to derive compact interpretable representations of flow fields. Representations for aerodynamic flow fields can be broadly segmented into linear and nonlinear techniques. Linear modal decomposition approaches are well-established in the aerodynamics literature and include proper orthogonal decomposition (POD), dynamic mode decomposition (DMD), Koopman analysis, global linear stability analysis, and resolvent analysis~\cite{taira2017modal, bui2004aerodynamic}. The POD and DMD methods are purely data-driven, applying matrix decomposition techniques, while Koopman, global linear stability, and resolvent analysis all are operator-based methods which linearize the governing operators. Linear modal methods have been used to study flows over a flat plate wing~\cite{ahuja2010feedback, ma2011reduced} and a bluff body~\cite{tu2013dynamic} at fixed Reynolds number as well as optimize airfoil shapes~\cite{bui2004aerodynamic}. Operator methods have been applied to studying flow instabilities over airfoils~\cite{he2017linear, rodriguez2011birth}. Among nonlinear modal decomposition techniques, spectral and neural approaches have garnered attention due to their ability to capture complex nonlinear distributions present in fluid flows~\cite{haller2016nonlinear,brunton2020machine,giral2026aerojepa}. A specific neural architecture, the autoencoder (AE), is naturally suited to representation learning tasks due to its bottleneck encoder-decoder architecture producing low-dimensional, \textit{i.e.}, \textit{latent space}, representations. The AE has been used to identify modes for a range of fluid systems such as flow over airfoils and cylinders~\cite{solera2024beta, eivazi2020deep, murata2020nonlinear, omata2019novel, fukagata2025compressing, frances2024toward, wang2024towards}. Variants of the AE method have been used to imbue structure on the latent space such as sparsity, latent alignment with aerodynamic coefficients, and dimensional energy-based ordering~\cite{solera2024beta, chen2025volumeanalysis, fukami2023grasping, fukami2020convolutional}. Neural methods have been shown to produce more compact representations than linear methods~\cite{solera2024beta}.

Separately, transfer learning techniques have been applied to data-driven studies of external aerodynamics to address data scarcity challenges. Researchers have investigated transferring airfoil and wing surface pressure fields, leveraging autoencoders to predict high-fidelity fields from their low-fidelity  counterpart ~\cite{nieto2025multi,shen2024application}. Similarly, transfer learning has been explored in the context of predicting airfoil pressure predictions from models trained on different Mach number and angle of attack regimes \cite{lian2026comparative}. Recently, researchers explored the use of Low-Rank Adaptation (LoRA) to transfer surface field prediction models between distinct automotive vehicle geometry classes \cite{keum2026adapting}. In addition, researchers have investigated transfer of deep learning surrogates, trained to predict 2D airfoil surface pressure fields, to prediction of 3D swept wing surface fields~\cite{runze2023transfer, yang2025rapid}. Lacking in thsi research doamin is an investigation of the transferability of compact flow representations derived from two-dimensional (2D) airfoils to complex three-dimensional (3D) aircraft shapes. Current representation learning approaches for aerodynamic fields therefore require hundreds to thousands of design-specific simulations, representing a severe computational barrier. Thus, whether representations learned from readily obtainable airfoil datasets can transfer to real-world design efforts in an accurate and data-efficient manner remains an open question that requires investigation.

\begin{figure}
\centering
\includegraphics[width=1.0\textwidth,  trim={3.5cm 1.5cm 3.75cm 2.9cm}, clip]{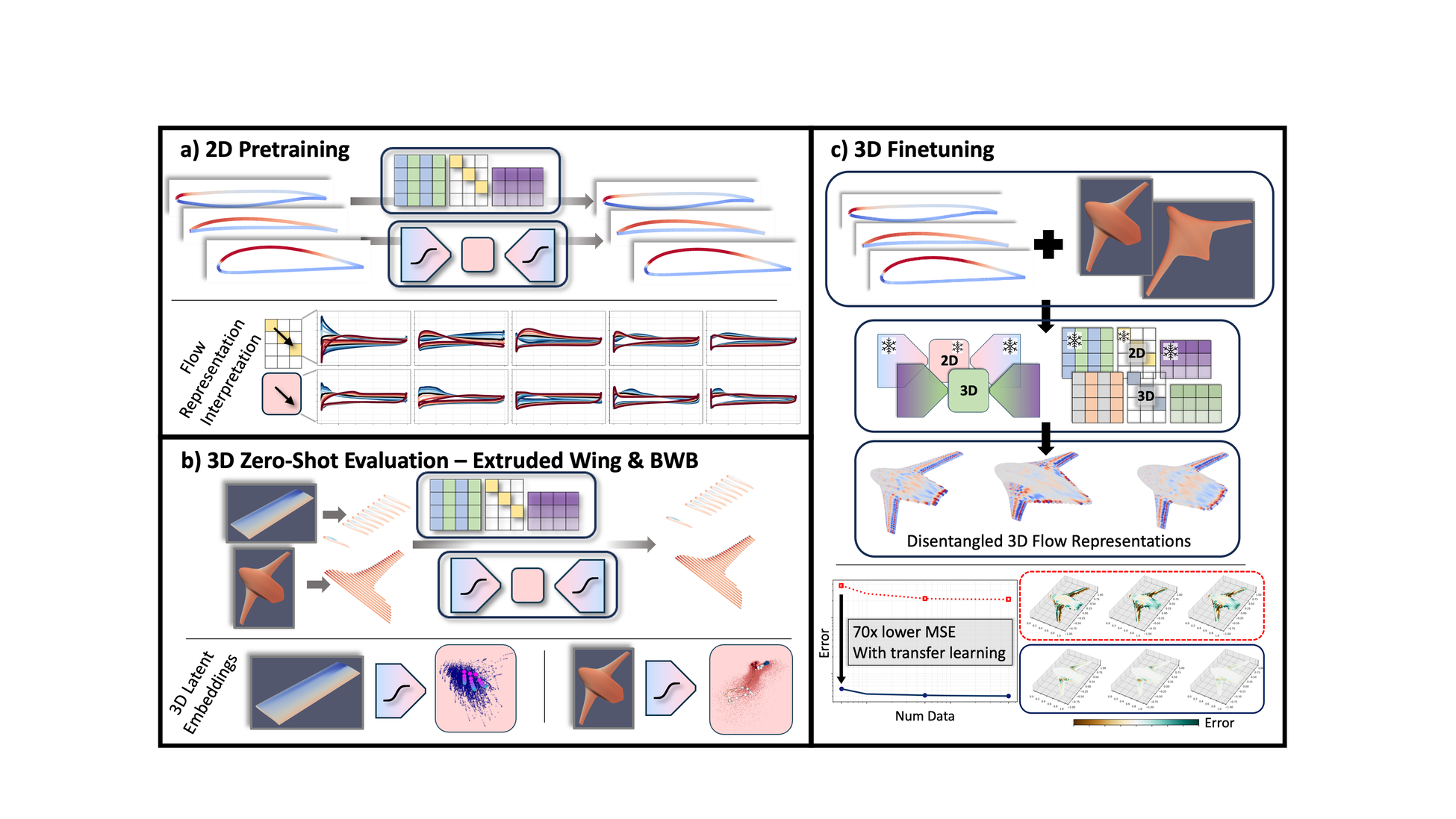}
\caption{Overview figure of this work depicting a) development of linear and nonlinear low dimensional representations from 2D datasets, b) the transfer of these representations to 3D geometries, and c) 3D fine-tuning of the flow representations.}
\label{fig:overview}
\end{figure}

To address these fundamental challenges, we learn compact representations of 2D airfoil surface fields using linear and nonlinear representation learning techniques and transfer these representations to complex 3D aircraft geometries. An overview of the approach is depicted in Fig.~\ref{fig:overview}. We find linear and nonlinear techniques produce compact (4-7$\times$ compression ratios) representations of 2D airfoil surface pressure fields, with the nonlinear method producing more accurate in-domain reconstructions than its linear counterpart at a fixed compression ratio. Furthermore, we demonstrate these 2D-based representations transfer in a {\it zero-shot} manner, accurately reconstructing surface pressure fields and lift and drag coefficients of complex 3D aircraft shapes. Strikingly, the linear method transfers {\it more effectively} than its nonlinear counterpart in both zero-shot and low-data regimes. Finally, the limited-data 3D regime is explored, where fine-tuning the low-dimensional representations using only five 3D cases reduces the representation error by over 5$\times$ relative to zero-shot transfer and by 1--2 orders of magnitude relative to models trained from scratch on equivalent data. The transfer learning techniques developed through this work allow scientists and engineers to perform modal analysis on a completely novel aircraft design with zero prior simulation or experimental data points. Previously, such analyses would have required hundreds to thousands of data samples. Thus transfer learning enables an extremely data-efficient route to interpreting flow fields, accelerating cutting-edge aircraft design efforts where little to no data exists.

The remainder of the paper is organized as follows. In \ref{sec:methodology-data} we present four aerodynamic datasets used for training and model evaluation. In \ref{sec:methodology-models} the linear and nonlinear representation learning models used in this study are reviewed. In \ref{sec:results-pressure-lvae} we present the results of applying the representation learning models to 2D airfoil datasets. In section \ref{sec:results-transfer} we assess the {\it zero-shot} transfer ability of the models to extruded 3D wing and blended wing body aircraft geometries, respectively. In \ref{sec:results-lvae-finetuning-bwb} we examine a composite representation learning model incorporating a 2D-based pretrained model and a 3D-based residual model trained on varying amounts data. In \ref{sec:conclusion} we summarize the key findings of this work.

\section{\label{sec:methodology}Data Generation and Machine Learning Methodology}

\subsection{\label{sec:methodology-data}Data Generation}

Surface pressure field representations are learned and evaluated across four datasets. Specifically, we generate two large datasets of 2D airfoil surface pressure fields. Additionally, a modestly sized 3D extruded-wing dataset, first presented by the authors in~\cite{diniz2026optiwing3d}, provides three-dimensional wing surface field data corresponding to an extruded cross-sectional geometry. The fourth dataset is an open source dataset comprised of blended wing body geometries and surface pressure fields~\cite{sung2025blendednet}. All datasets were generated using computational fluid dynamics (CFD) solvers that solve the Reynolds-Averaged Navier-Stokes equations. The Spalart-Allmaras turbulence model was used as a closure model for the system of equations.

\subsubsection{\label{sec:methodology-data-engibench}Two-Dimensional Aerodynamic Dataset}

Two 2D datasets were comprised of 93,500 and 55,000 airfoil simulations respectively. Simulations were performed using the EngiBench data generation library~\cite{felten2025engibench}. EngiBench orchestrates the CFD simulations in parallel on high-performance computing (HPC) resources, and the MACH-Aero solver suite is used to simulate each individual case \footnote{MDO Lab, \textit{MACH-Aero}, \url{https://github.com/mdolab/MACH-Aero.git}}. MACH-Aero provides pre-processing, geometry parameterization, and fluid solver routines required to perform the analysis~\cite{Mader2020a, Hajdik2023, Secco2021}. 

The two datasets are termed 2D Dataset 1 and 2D Dataset 2. Two datasets were generated to align the flow and angle of attack regimes with the two distinct target 3D datasets. The dataset geometries were created by selecting 935 airfoil geometries from the UIUC database~\cite{selig1996uiuc} as well as shape optimized variants~\cite{diniz2024optimizing}. For each geometry in 2D Dataset 1, approximately 50 Mach number, Reynolds number, and angle of attack tuples are sampled using Latin Hypercube Sampling (LHS). For 2D Dataset 2 approximately 30 Mach number, Reynolds number, and angle of attack tuples are sampled for each geometry. The sampling regime for the two datasets is distinct with exact values provided in Table \ref{tab:dataset_ranges}. Each geometry, flow parameter, and angle of attack data combination was then simulated using the CFD solver. The resulting comprehensive 2D datasets cover a wide swathe of geometric design space, flow conditions, and angle of attack values. 

\subsubsection{\label{sec:methodology-data-wing}OptiWing3D Dataset}

The OptiWing3D dataset consists of optimized, variable-cross section, extruded-wing geometries. The dataset consists of  935 airfoil geometries from the UIUC database~\cite{selig1996uiuc}, which were extruded to 2.5 times the chord length in the spanwise direction and then optimized. Although the dataset consists of initial and optimized wing pairs, the current paper uses only the initial-wing simulations from that dataset. Each extruded wing geometry is paired with a randomly sampled Mach and Reynolds number from the same parameter range as Dataset 1\textemdash these values are shown in Table \ref{tab:dataset_ranges}. The angle of attack for each case is held fixed at 2.5$^\circ$. These cases were then simulated using the MACH-Aero framework, which features the differentiable ADflow solver\cite{Mader2020a}. For a complete description of the dataset, we refer readers to the work of Diniz et. al.\cite{diniz2026optiwing3d}.

Note that the airfoil/cross-section geometry and flow parameter ranges are shared between 2D Dataset 1 and the 3D extruded wing dataset. However, flow conditions were sampled independently in each and therefore no combination of cross-section geometry, flow conditions, and angle of attack from one dataset appears in the other.

\subsubsection{\label{sec:methodology-data-blendedbody}Three-Dimensional Blended Wing Body Dataset}

The fourth aerodynamic dataset is comprised of blended wing body (BWB) 3D geometries. The BWB geometry smoothly blends the fuselage and wing of the aircraft into a single lifting surface. Due to enhanced lifting characteristics and reduced wetted area, blended wings have the potential to improve aerodynamic efficiency and enhance fuel economy versus traditional aircraft designs~\cite{qin2004aerodynamic}. However, due to their unique shape BWB aircraft exhibit significant geometric complexity, incorporating chord, span, sweep, and taper variations. The BWB shape inherently introduces three-dimensional flow structures, in particular strong spanwise flow. Thus the blended wing data presents a compelling and challenging 2D to 3D transfer and fine-tuning problem.

The BWB dataset was made publicly available through the Harvard Dataverse at~\cite{DVN/VJT9EP_2025} and used to develop the BlendedNet model~\cite{sung2025blendednet}. The geometric design is controlled by eight independent parameters defining chord length, span width, and sweep angle. A consistent cross-sectional airfoil shape was used. The geometric parameters were sampled using LHS to generate a dataset of 999 unique blended wing body shapes. Each blended wing geometry was paired with 10 LHS flow conditions (Mach number, Reynolds number, and angle of attack) drawn from the flow parameter range shown in Table \ref{tab:dataset_ranges}. Note that the BWB dataset shares the same flow and angle of attack regime as 2D Dataset 2. The BWB dataset is partitioned by geometry into 80\% train and 20\% validate subsets. Each geometry is unique to a given split and therefore does not appear within the other split. For a complete description of the geometry generation and partitioning process refer to the work of Sung et. al.\cite{sung2025blendednet}.

\begin{table*}
\renewcommand{\arraystretch}{1.}
\caption{\label{tab:dataset_ranges} Aerodynamics Datasets and Associated Parameter Ranges}
\begin{tabular}{p{5.5cm} c c c c}
\toprule
Dataset & \makecell{Number\\of Cases} & \makecell{Mach\\Number} & \makecell{Reynolds\\Number ($\times10^6$)} & \makecell{Angle of\\Attack {[deg]}}\\ 
\midrule
2D Dataset 1 (UIUC\cite{selig1996uiuc} + Opt\cite{diniz2024optimizing}) & 93,500 & 0.5 -- 0.9 & 1.0 -- 20.0 & 0 -- 20 \\[4pt] 
2D Dataset 2 (UIUC\cite{selig1996uiuc} + Opt\cite{diniz2024optimizing}) & 55,000 & 0.05 -- 0.5 & 0.1 -- 200.0 & -10 -- 20 \\[4pt] 
Extruded Wings (UIUC\cite{selig1996uiuc}) & 935 & 0.5 -- 0.9 & 1.0 -- 20.0  & 2.5 \\[4pt]
Blended Wings (BlendedNet\cite{sung2025blendednet})  & 10,000 & 0.05 -- 0.5 & 0.1 -- 200.0  & -10 -- 20 \\
\bottomrule
\end{tabular}
\end{table*}

\subsection{\label{sec:methodology-models}Representation Learning Models}

This study uses two representation learning model classes; a linear technique based on matrix decomposition, and a nonlinear neural network-based approach. Representation learning, a subfield of machine learning, is closely related to dimensionality reduction techniques from statistics and modal analysis techniques from fluid dynamics. These models provide informative and compact representations for aerodynamic surface fields.

\subsubsection{\label{sec:model-pca}Proper Orthogonal Decomposition}

The POD method is a technique for decomposing a flow field into orthogonal components, or {\it modes}. The decomposition is based on linear projection and reconstruction operations, to and from a low dimensional subspace, such that the error of the reconstructed flow fields minimizes a mean square error. As this is a linear technique, the projection and reconstruction are matrix operations, readily learned using linear algebra libraries. The POD approach is closely related to singular value decomposition (SVD) and thus inherits the associated eigenspectrum ordering characteristics. For surface pressure fields, the learned POD modes are thus ordered by the force magnitude contained within each mode. Truncating the POD modes to a fixed dimensionality ensures that the squared reconstruction error is minimized over all possible linear projections to the same dimensional subspace and that the maximum amount of total force is captured within that subspace. For a complete review of the POD method for modal analysis of fluid flows the reader is referred to~\cite{taira2017modal, bui2004aerodynamic}.

In this work the POD model is trained on both 2D airfoil datasets, as well as a subset of the three dimensional data. For all training scenarios, the pressure profiles are either 2D airfoil slices or 2D cross-sectional slices of 3D data (oriented normal to the spanwise direction) and are represented as 192 dimensional vectors. Each vector is constructed by sampling points starting at the trailing edge, traversing the upper surface to the leading edge, and returning along the lower surface to the trailing edge. To ensure a consistent pressure data representation we use a radial basis function to interpolate data from disparate sources to this consistent 192-dimensional representation. For complete information on the sampling and interpolation procedure refer to appendix \ref{app:data_processing}. The resulting data matrix has the dimensions $N\times192$ with $N$ corresponding to the number of cases for a given dataset. The Scikit-learn~\cite{scikit-learn} Python library is used to perform the exact full SVD operation. Subsequently, the number of singular values\textemdash alternatively referred to as components or modes\textemdash are truncated to achieve a low-dimensional representation.

Once the POD model has been fit to a dataset, the decomposition is evaluated on reconstructing the train, validate, and test splits of all datasets. When comparing to the nonlinear model, the number of POD modes is chosen to either match the nonlinear model dimensionality or a certain reconstruction accuracy threshold. When matching a certain reconstruction accuracy threshold, the accuracy on the validation data split is used.

\subsubsection{\label{sec:model-lvae}Least Volume Autoencoder}

The Least Volume autoencoder\cite{chen2025volumeanalysis} (LVAE) operates on the basis of the {\it manifold hypothesis}. According to the manifold hypothesis~\cite{fefferman2016testing, bengio2013representation} high-dimensional aerodynamic data, represented, \textit{e.g.}, as cell, node, or point cloud data, lie on a lower dimensional nonlinear manifold.  Like other AEs, the LVAE method learns to encode high-dimensional data, \textit{e.g.}, pressure fields, to a latent representation and subsequently reconstruct, or decode, it back to the original data space. The LVAE method was developed to automatically discover the intrinsic, {\it i.e., least volume}, dimension of the data. Thus, in principle, the LVAE learns the most compact representation of a data distribution which preserves the distribution structure. The process of learning the intrinsic dimension \textit{and} data reconstruction is performed simultaneously during the training phase. 

The data reconstruction task is achieved by minimizing a standard mean squared error (MSE) loss. The least volume dimension is identified through a combination of a volume penalty applied to the latent space, a Lipschitz-constraint placed on the decoder, and a dynamic pruning algorithm to remove dimensions with nearly zero data variation. The volume penalty morphs and compresses the latent data distribution to align with a low dimensional subspace. The Lipschitz normalization prevents isotropic shrinkage of the latent space, enforcing a smooth mapping between the latent space and raw data representations of the pressure. Finally, the dynamic pruning algorithm freezes latent dimensions with {\it approximately zero} standard deviation magnitude. These frozen or \textit{inactive} dimensions are replaced by a fixed constant mean value. As the latent space evolves during training the pruning process is iteratively repeated until the model converges to a compact latent representation. The standard deviation of the latent dimensions provides an ordering to the LVAE model conceptually similar to that of the POD approach. A further connection between the nonlinear and linear approaches is due to the fact that the SVD algorithm underlying POD is a limiting case of the LVAE model when the encoder and decoder networks are restricted to linear operations. The LVAE optimization process is multi-objective, balancing compactness with accurate reconstruction of the aerodynamic pressure fields. A complete description of the LVAE method is detailed in the original work by Chen \textit{et al.}\cite{chen2025volumeanalysis}. 

In this work, the LVAE model is trained on both 2D airfoil datasets, as well as a subset of the 3D data. For all cases, the pressure profiles are either 2D airfoil slices or 2D cross-sectional slices of 3D data (oriented normal to the spanwise direction) and are represented in an identical manner as described in the preceding POD section. We use a one-dimensional convolutional neural network (CNN) architecture for both our encoder and decoder in the LVAE architecture. In addition, the decoder is spectrally normalized to globally constrain the Lipschitz constant. Validation-based checkpointing is used to store the model at the epoch for which the validation loss is minimized or the minimum latent dimension is reached. The checkpointed model is used as the {\it final} model for subsequent test set evaluation and inference tasks. Complete details of the LVAE architecture and training strategy is presented in App. \S\ref{app:lvae_arch}.

Both the POD and LVAE models are assessed by 1) the representation dimensionality (\textit{i.e.}, number of singular values or least volume dimension), 2) the MSE reconstruction accuracy of the coefficient of pressure ($C_p$), and 3) the ability to preserve lift ($C_L$) and drag ($C_D$) coefficient values quantified using the mean absolute percent error metric (MAPE).

\section{\label{sec:results}Results and Discussion}

\subsection{\label{sec:results-pressure-lvae}Learned Representations of 2D Pressure Data}

The metrics for the POD and LVAE models trained on the 2D datasets are presented in Table \ref{tab:lvae_training_results}. The LVAE model obtains a latent dimension of 28 and 41 for 2D Datasets 1 and 2 respectively, reflecting a compression ratio of approximately 6.8$\times$ and 4.7$\times$. Furthermore, the reconstruction MSE for the test set is $3.0\times10^{-5}$ and $5.5\times10^{-4}$, respectively. The relatively higher MSE and latent embedding dimension for 2D Dataset 2 is attributed to the larger Reynolds number and angle of attack regime (which generates a more complex flow/data distribution) coupled with the smaller dataset size. The training curves for the LVAE models are presented in App. \S\ref{app:lvae_training_curves}. For each dataset, two POD models are obtained by retaining different numbers of the singular values. The first is simply obtained by retaining only the same number of singular values as the LVAE latent dimension. The second is obtained by identifying the number of singular values required to match the validation accuracy of the LVAE model. For both 2D datasets, the LVAE model has higher reconstruction accuracy than the POD model for an equivalent latent dimension. Furthermore, the POD approach requires 37 and 13 more dimensions, for 2D Datasets 1 and 2, respectively, to match the LVAE performance. The LVAE model outperforms POD on the task of learning compact representations as it is able to learn a nonlinear data manifold whereas POD requires more linear subspaces to capture the same amount of data variation.

\begin{table*}
\renewcommand{\arraystretch}{1.}
\centering
\caption{\label{tab:lvae_training_results} POD and LVAE Model Error Metrics Evaluated on 2D Data Test Sets}
\begin{tabular*}{\textwidth}{@{\extracolsep{\fill}} p{3.0cm} c c c c c @{}}
\toprule
Dataset & Model & \makecell{$C_p$ MSE} & \makecell{$C_L$ MAPE} & \makecell{$C_D$ MAPE} & Dimension \\
\midrule
\multirow{3}{*}{2D Dataset 1} & LVAE & $3.01\times10^{-5}$ & 0.43 & 0.98 & 28 \\[4pt]
\arrayrulecolor{lightgray}\cmidrule{2-6}\arrayrulecolor{black}
 & POD & $8.53\times10^{-4}$ & 0.11 & 0.65 & 28 \\[4pt]
 & POD & $3.08\times10^{-5}$ & 0.005 & 0.03 & 65 \\[4pt]
\midrule
\multirow{3}{*}{2D Dataset 2} & LVAE & $5.53\times10^{-4}$ & 1.17 & 8.99 & 41 \\[4pt]
\arrayrulecolor{lightgray}\cmidrule{2-6}\arrayrulecolor{black}
 & POD & $1.36\times10^{-3}$ & 0.24 & 2.73 & 41 \\[4pt]
 & POD & $5.65\times10^{-4}$ & 0.08 & 0.74 & 54 \\
\bottomrule
\end{tabular*}
\end{table*}

The pressure-induced lift and drag coefficients are critical performance metrics of an airfoil, indicating the force generated normal, and tangential, to the incoming flow direction. Both pressure-induced lift and drag are integrated quantities of the pressure profiles and can be calculated for the ground-truth and reconstructed pressure distributions. To avoid division by small numbers the reconstructed lift and drag MAPE are evaluated on airfoil cases whose ground-truth lift and drag values exceed $10^{-2}$ and $10^{-3}$ respectively (this threshold retains $>99$\% of the total data). The MAPE of these quantities is reported in Table \ref{tab:lvae_training_results}. Both the POD and LVAE models preserve lift and drag to within 1\% and 9\% for lift and drag respectively. However, despite the better $C_p$ reconstruction accuracy of LVAE, POD preserves the $C_L$ and $C_D$ coefficients more accurately. This is attributed to the fact that lift and drag are surface integral quantities and therefore pointwise error can be canceled out when calculating this integrated quantity. Quantitatively, the LVAE residuals exhibit a 5$\times$ tighter standard deviation than POD, but a mean residual of $-1\times10^{-4}$ versus $4\times10^{-7}$ for POD; thus the LVAE bias accrues coherently during the surface integration, producing the larger lift and drag errors despite the lower pointwise scatter. 

\begin{figure}[ht!]
\centering
\includegraphics[width=1.0\textwidth]{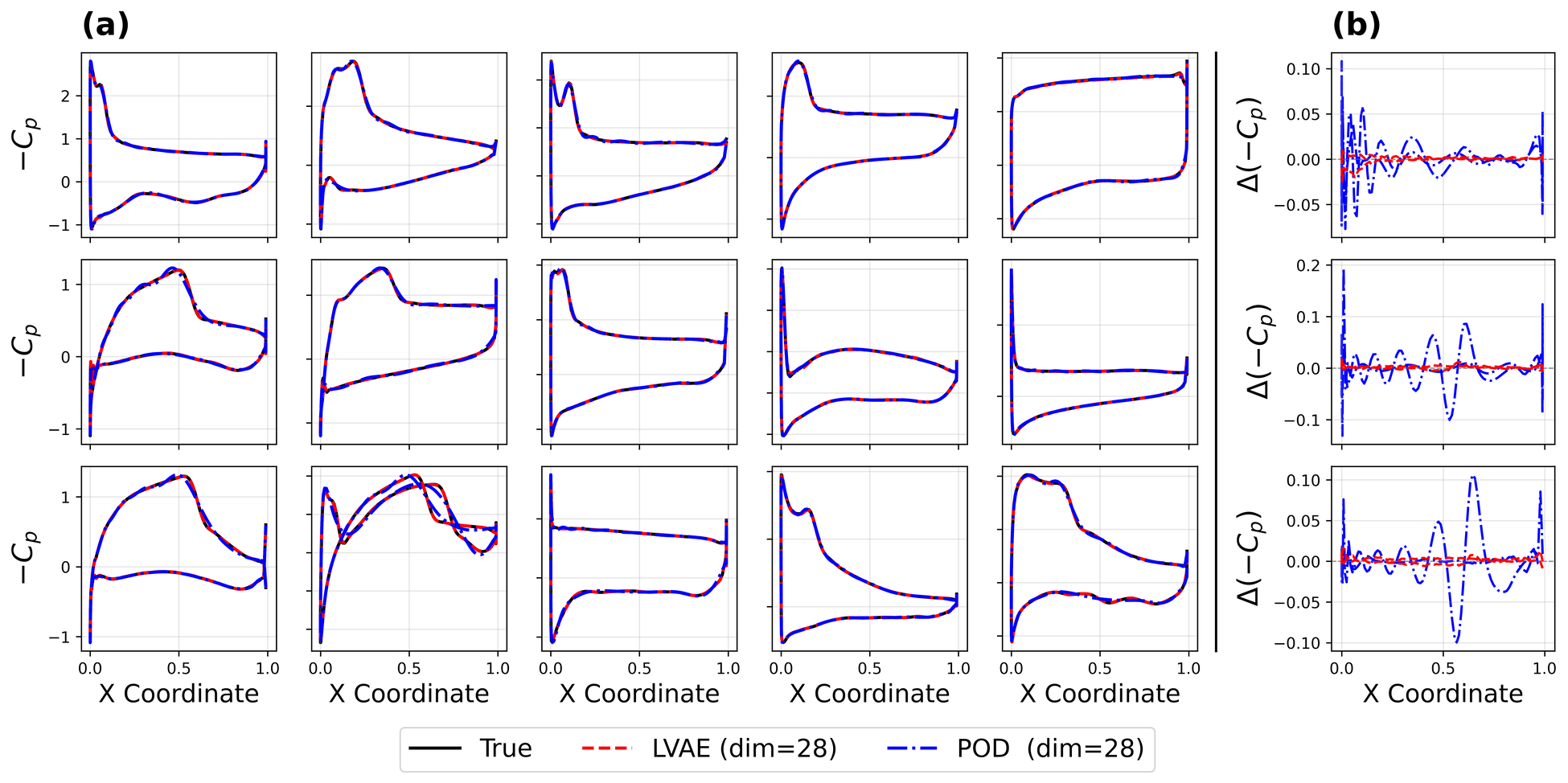}
\caption{POD and LVAE (a) reconstructions and (b) residuals for test set airfoil pressure data.}
\label{fig:reconstruct_engibench_test}
\end{figure}

To obtain a more granular view of the reconstructed airfoil pressure profiles we plot a random sample of 2D Dataset 1 test set airfoil pressure profiles versus the airfoil $x$-coordinate in Fig.~\ref{fig:reconstruct_engibench_test}. From the reconstruction plots it is clear that despite a wide variety of pressure curve shapes, both the POD and LVAE models are capable of high-accuracy reconstructions from a 28-dimensional latent space. The residual plots indicate that the pointwise reconstruction error is significantly larger for the POD model when the dimensionality matches that of LVAE. This observation corroborates that the lower POD lift and drag MAPE values arise from cancellation of these larger, but unbiased, pointwise errors during surface integration.

\begin{figure}[ht!]
\centering
\includegraphics[width=1.0\textwidth]{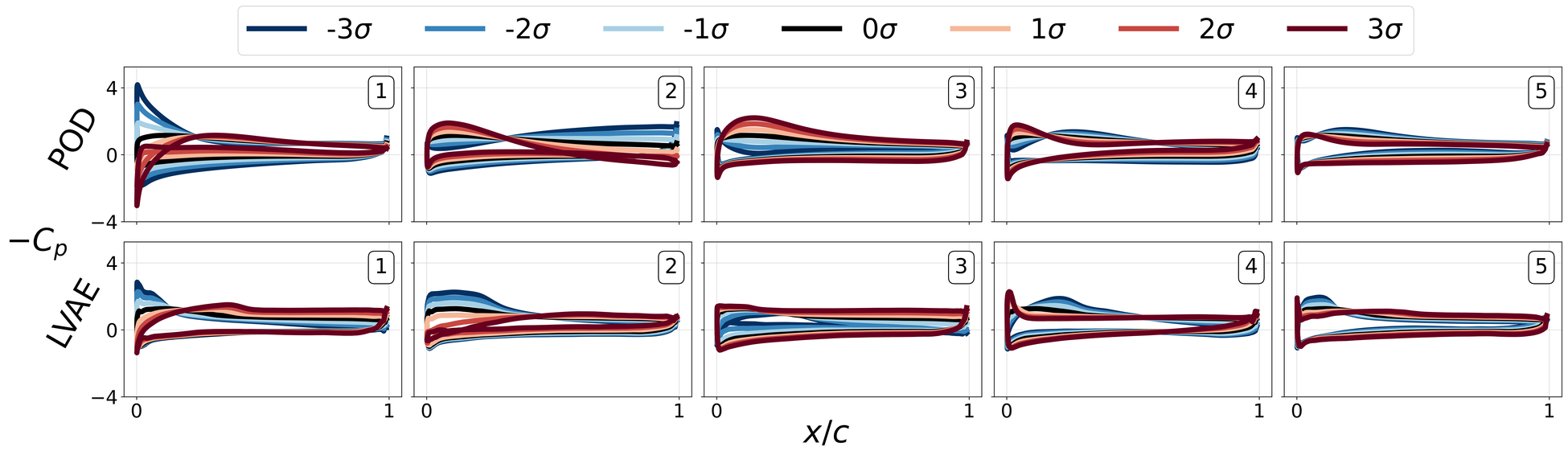}
\caption{POD (top row) and LVAE (bottom row) reconstructed pressure fields corresponding to latent interpolation about the mean latent embedding for the five most important latent dimensions of 2D Dataset 1.}
\label{fig:latent_interpolation}
\end{figure}

A key benefit of compact representation is the ability to interrogate the dominant flow structures of a fluid system. We perform this analysis by investigating how variation in latent dimension influences the decoded pressure profiles. Specifically, we interpolate through all latent dimensions, of the POD and LVAE models, and visualize this effect on the reconstructed pressure data. For each latent dimension $i$, the interpolated latent vector is generated as $z_{interp, i} = z_{mean} + j*\sigma_i \bf{1}_i$ where $z_{mean}$ is the mean embedding vector of all training data, $\sigma_i$ is the standard deviation of the latent embeddings for dimension $i$, and $\bf{1}_i$ is the indicator vector which has a value 1 for latent dimension $i$ and zeros for all other dimensions. The interpolation about the mean embedding vector is performed individually for each latent dimension, recall that both POD and LVAE have latent dimensions ordered by the magnitude of pressure MSE improvement. Each interpolated latent vector is reconstructed to obtain the associated pressure profiles. The reconstructed surface pressure {\it modes} for the POD and LVAE models are shown in Fig.~\ref{fig:latent_interpolation}. The full set of latent interpolation plots for the POD and LVAE models trained on 2D Datasets 1 and 2 are shown in App. \S\ref{app:latent_interpolation}. 

In Fig.~\ref{fig:latent_interpolation}, we observe the reconstructed pressure field from the mean latent vector shown as a black line. For interpolation within each latent dimension, a set of deviations from this mean curve are plotted, with each colored curve corresponding to interpolating 1, 2, or 3 standard deviations in the negative and positive directions. The most important latent dimensions (first column) for POD and LVAE produce large variations in the pressure profile near the upper surface leading edge. However the first POD mode also significantly varies the lower surface pressure distribution while the LVAE mode produces minimal lower surface distribution. The second POD mode significantly shifts the pressure distribution along the upper surface from roughly the quarter-chord to the trailing edge, in contrast the LVAE mode produces mainly upper surface leading edge to quarter-chord variations. The third POD mode primarily shifts the pressure distribution along the entire upper surface, in contrast the LVAE mode reduces the pressure magnitude along the upper and lower surfaces. The fourth and fifth modes for POD and LVAE produce more localized pressure variations exhibiting node and anti-node like behavior about the mean curve. This trend extends to less important latent dimensions which map to highly localized and small magnitude variations of the pressure distribution (see appendix \ref{app:latent_interpolation}).

We have demonstrated that the representation learning models provide a compact (4.7 - 6.8$\times$) and highly accurate latent representation. These models can capture aerodynamic pressure fields arising from 2D simulations spanning a broad range of airfoil geometries, flow conditions, and angles of attack. The nonlinear LVAE model outperforms POD on the surface field reconstruction task, while POD more accurately maintains lift and drag coefficients. Inspection of the low-dimensional representations reveal the most critical flow patterns within the 2D datasets. Next we assess the {\it zero-shot} transfer capability on two sets of three-dimensional aerodynamic shapes. First, for a quasi-2D extruded wing geometry and, second, for complex blended wing body aircraft shapes.
\subsection{\label{sec:results-transfer}Zero-Shot Representation Transfer}

\subsubsection{\label{sec:results-transfer-wing}Extruded Wing}

The POD and LVAE models pretrained on 2D Dataset 1 are applied to reconstruction of 3D wing cross-sectional flow fields taken from the extruded wing dataset described in \S\ref{sec:methodology-data-wing}. In particular, we encode two-dimensional slices of the wing where each slice is taken normal to the span-wise coordinate. Nine unique span-wise locations from 4\% to 89\% of the total span width are used in this analysis. This ensures locations are sampled both near the wing root, where the flow will appear more {\it 2D-like}, and closer to the wing tip where 3D flow effects are strongest. In Table~\ref{tab:lvae_wing} we report the MSE for the total reconstruction error of all wing cases and all cross-sections. Furthermore, to isolate the span effect we report the reconstruction MSE for all 3D wings at three unique cross section locations -- those nearest the root, mid-span, and wing-tip locations.

The $C_p$ reconstruction, and reconstructed lift and drag coefficient, error for the 3D wing is larger in magnitude than for the 2D airfoils. The relatively higher error is unsurprising given that the data is generated from three dimensional simulations containing flow in the span direction, wing tip vortices, and other intrinsically three-dimensional fluid structures. Despite the three dimensional nature of the flow it is notable how adeptly the POD and LVAE models, trained purely on 2D data, transfer to this three dimensional system. Furthermore, note a positive correlation between the error metrics and the spanwise coordinate. The error is smallest near the root, where the flow is quasi two-dimensional, and increases near the tip where more three-dimensional flow effects are present. While the reconstruction performance is roughly equivalent, $6-7.5\times10^{-4}$ $C_p$ MSE for the POD and LVAE models, POD significantly outperforms LVAE on maintaining lift and drag coefficients. Indeed, the overall lift and drag MAPE is 0.1\% and 0.68\% for POD versus 1.47\% and 4.57\% for LVAE. The higher lift and drag performance of the POD model is attributed to the fact that the residuals are evenly distributed about zero while the LVAE residuals are biased to negative values. Thus the pointwise errors of the POD model tend to cancel whereas the bias of the LVAE predictions leads to higher lift and drag errors. 

\begin{table*}
\renewcommand{\arraystretch}{1.}
\centering
\caption{\label{tab:lvae_wing} POD and LVAE Extruded Wing Representation Transfer Metrics at All Cross-Sections and at Root (4\%), Mid-Span (47\%), and Tip (89\%) Locations}
\begin{tabular*}{\textwidth}{@{\extracolsep{\fill}} p{3.0cm} c c c c c c @{}}
\toprule
& \multicolumn{2}{c}{\makecell{$C_p$ MSE $[\times10^{-4}]$}} & \multicolumn{2}{c}{\makecell{$C_L$ MAPE {[\%]}}} & \multicolumn{2}{c}{\makecell{$C_D$ MAPE {[\%]}}} \\
\cmidrule(lr){2-3} \cmidrule(lr){4-5} \cmidrule(lr){6-7}
Cross Section & POD & LVAE & POD & LVAE & POD & LVAE \\
\midrule
3D -- All      & 6.37 & 6.57 & 0.10 & 1.47 & 0.68 & 4.57 \\[4pt]
3D -- Root     & 5.92 & 6.34 & 0.09 & 1.21 & 0.67 & 4.43 \\[4pt]
3D -- Mid-Span & 6.19 & 6.47 & 0.09 & 1.34 & 0.67 & 4.54 \\[4pt]
3D -- Tip      & 7.49 & 7.37 & 0.17 & 2.34 & 0.72 & 4.70 \\
\bottomrule
\end{tabular*}
\end{table*}

\begin{figure}
\centering
\includegraphics[width=1.0\textwidth]{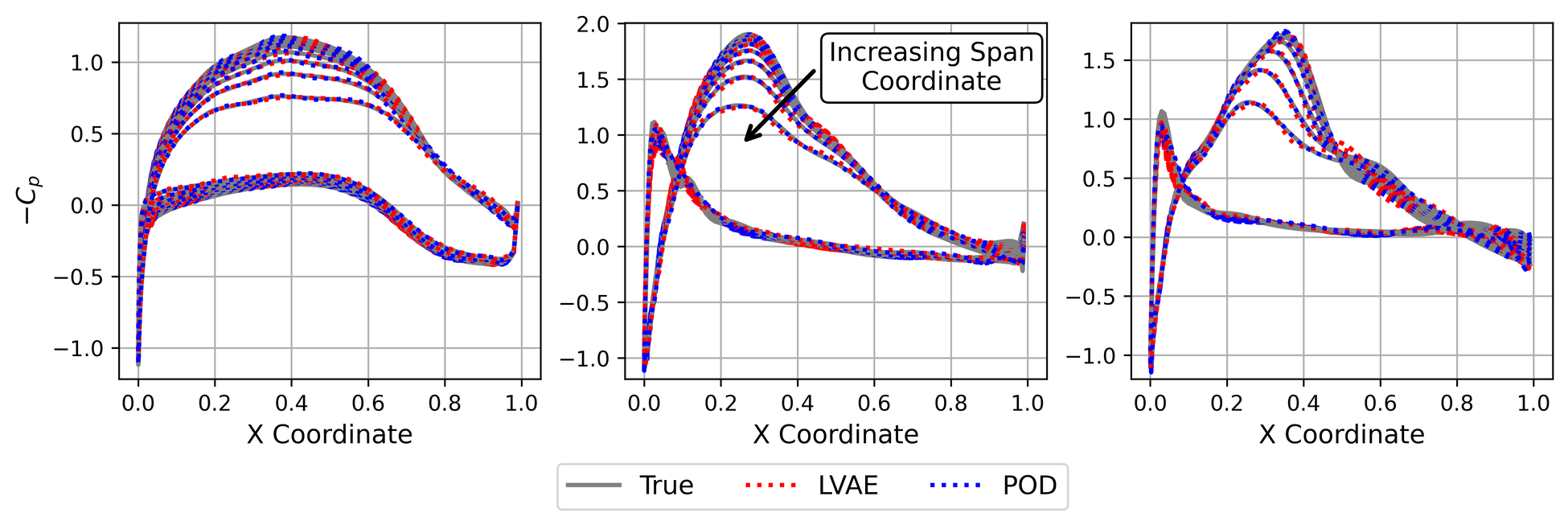}
\caption{The ground-truth (black), POD (blue), and LVAE (red) pressure fields of three wings across all nine cross sections. Each plot corresponds to a unique wing case, \textit{i.e.}, geometry and flow conditions. The direction of increasing span coordinate, moving away from the root and toward the wing tip, is indicated in the central plot with a black arrow.}
\label{fig:wing_reconstruction}
\end{figure}

It is beneficial to extend our analysis to the set of cross-sections comprising an entire wing, as this collection of spans represents a fully three-dimensional aerodynamic object. To understand how our models perform at this wing-level task, we visualize reconstructed pressure fields for three randomly sampled wings in Fig.~\ref{fig:wing_reconstruction}. The pressure fields for the wing cross-sections clearly exhibit a spanwise variation with the upper surface (larger negative $C_p$) varying more strongly along the span of the wing than the bottom surface. Both POD and LVAE models capture the spanwise variation, reproducing the decrease in $C_p$ magnitude near the wing tip. The reconstructed fields are not perfect; in particular rapid pressure fluctuations are not always captured. For example in the third plot, the top surface variations past half a chord length are not precisely reconstructed. Despite this deficiency, the models do maintain the spanwise ordering of the original data at nearly all chord locations. Namely, at a given chord length if the original pressure is increasing or decreasing along the span then this increase or decrease is maintained in the reconstructed data. Despite the reduction in accuracy when transferring the 2D  pretrained representations models to 3D wings, the reconstruction-based lift and drag (integrated over the wing cross-sections using the trapezoid method) are 0.1\% and 0.7\% respectively for POD and 1.4\% and 4.3\% for LVAE. This indicates the compact low-dimensional representations capture key performance metrics for the entire 3D aerodynamic shape.

\begin{figure}[ht!]
\centering
\includegraphics[width=1.0\textwidth,  trim={3.5cm 0cm 2.5cm 0cm}, clip]{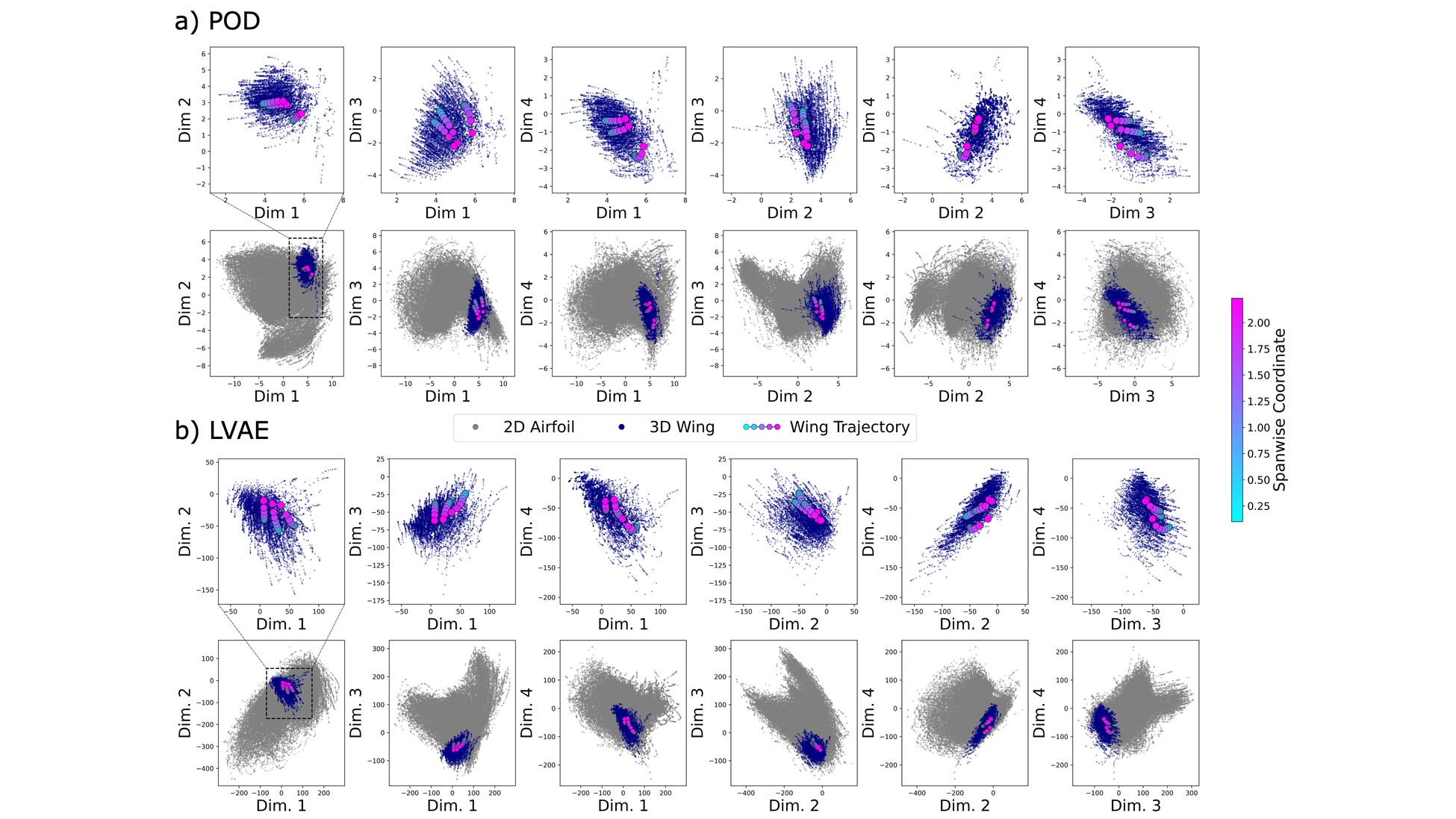}
\caption{Pairwise latent embeddings of 2D airfoil (grey) and 3D wing (blue) data for (a) POD and (b) LVAE. Upper rows magnify the 3D subregion, showing three wing cases with embeddings colored by span coordinate from root (cyan) to tip (pink).}
\label{fig:2D_3D_latent_embeddings}
\end{figure}

Visualizing the latent embeddings of 2D airfoils and 3D wing cross-sections in Fig.~\ref{fig:2D_3D_latent_embeddings} provides insight into why the POD and LVAE models transfer accurately to the 3D wings. Since the latent space is high-dimensional, only the pair-wise embeddings of the four most important latent dimensions are visualized. The 3D wing data (colored points) are embedded within a localized region of the entire 2D latent embedding distribution (grey points), shown in the lower row of the POD and LVAE subplots in Fig.~\ref{fig:2D_3D_latent_embeddings}. The in-distribution embedding, in conjunction with the relatively high accuracy of the 3D wing reconstructions, indicates that the 3D cross-section pressure distributions resemble those seen within the 2D airfoil training set. Thus, despite the significant spanwise variation of 3D flow shapes, the surface field at any particular cross-section appears similar to those in the 2D dataset. Recall that aside from a fixed angle of attack, all geometric and flow condition parameters share a common distribution between the 2D airfoil and 3D wing datasets. Therefore, we surmise that the localization of the 3D embedding distribution is a result of a fixed angle of attack ($\alpha=2.5^{\circ}$), indicating the critical role this parameter plays in the shapes of the surface pressure fields and thus the latent space topology. In the upper rows of Fig.~\ref{fig:2D_3D_latent_embeddings}, we inspect the subregion occupied by the 3D wing embeddings and visualize how wing cross-section embeddings vary with respect to span for three 3D wing cases. We observe that varying along the span coordinate of a wing corresponds to smooth variation in the latent space. Furthermore, embeddings near the wing root (cyan) are more proximal in the latent space than embeddings near the tip (pink) due to the increasing magnitude of 3D flow effects at the wing tip leading to larger distortion of the surface fields.

\subsubsection{\label{sec:results-transfer-bwb}Blended Wing Body}

Blended wing aircraft are aerodynamic geometries incorporating variations in span, sweep, and taper. These 3D geometric variations lead to complex pressure distributions and unique lift, drag, and pitching moment characteristics~\cite{re2005longitudinal, roman2000aerodynamic, carter2006designing, sung2025blendednet}. Given this surface field complexity, the transfer of the POD and LVAE-based flow representations learned from 2D is a more challenging task than transfer to the extruded wing geometries. We apply the models pretrained on 2D Dataset 2. This choice is made to ensure that the transferred model is pretrained on an identical distribution of flow and angle of attack conditions as the BWB dataset. This isolates the geometric component of transfer from any flow-condition distribution shift, \textit{e.g.}, transfer from high to low Mach number regimes.

\begin{figure}
\centering
\includegraphics[width=1.0\textwidth]{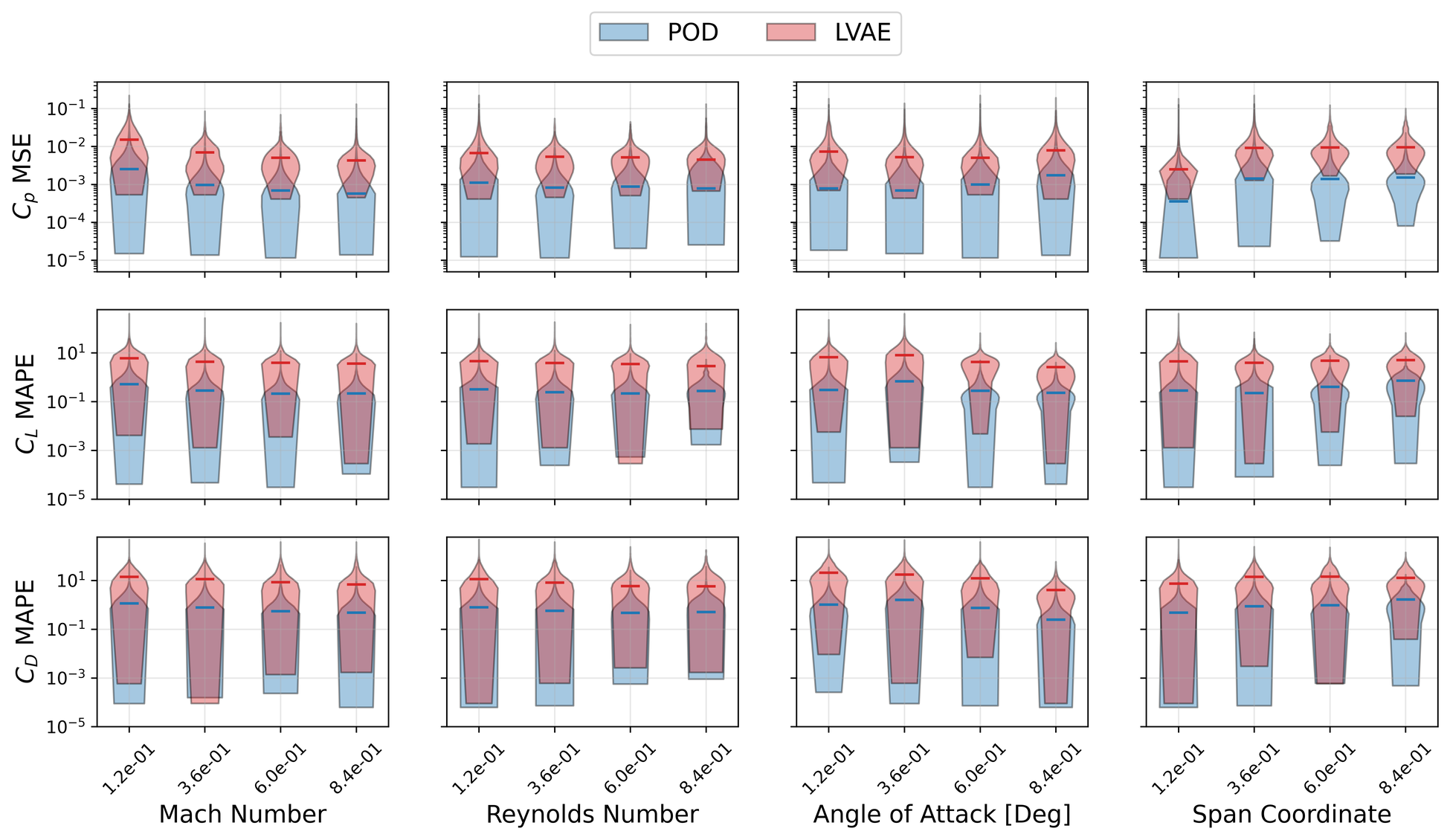}
\caption{BWB reconstruction error distributions for MSE (top), lift MAPE (middle), and drag MAPE (bottom) versus Mach number, Reynolds number, angle of attack, and span coordinate ({\it left to right}). Errors are binned into four uniform intervals per variable; horizontal bars indicate medians.}
\label{fig:bwb_reconstruction_dists}
\end{figure}

The BWB dataset reconstruction MSE is plotted against Mach number, Reynolds number, angle of attack, and span coordinate in the top row of Fig.~\ref{fig:bwb_reconstruction_dists}. For the POD model the mean and median MSE is $2.6\times10^{-3}$ and $1.2\times10^{-3}$, respectively, and $1.1\times10^{-2}$ and $6.2\times10^{-3}$, respectively, for LVAE. Furthermore, both models exhibit skew toward lower errors. Strikingly, the POD model has significantly lower MSE, indicating that the linear learned representations transfer more effectively than their nonlinear counterpart, with both models using an equal number of dimensions. As shown in Fig.~\ref{fig:bwb_reconstruction_dists}, the POD model consistently produces median $C_p$ reconstructions 5 to 7$\times$ more accurate than LVAE across all Mach number, Reynolds number, angle of attack, and span coordinate regimes. Both models exhibit relatively higher magnitude of MSE, compared to the 2D or extruded 3D wing scenarios. These higher errors are attributed to the geometric complexity of BWB aircraft shapes generating flow patterns not present in the 2D dataset. The $C_p$ MSE decreases as Mach number increases, indicating that BWB pressure distributions are more dissimilar to 2D at low Mach numbers. The MSE variation with respect to Reynolds number and angle of attack is relatively weak. Finally, a strong increase in error with increasing span coordinate is evident. This aligns with intuition as well as behavior observed in the 3D extruded wing where cross-sections closer to the wing tip exhibit higher magnitude 3D flow effects. Note that for BWB the increase in error with span plateaus near the mid-span region corresponding to the location where the BWB cabin merges with the wing portion. This transition zone leads to an onset of spanwise flow features causing the pressure distributions to deviate more strongly from those present in the 2D dataset.

The ability of flow representations to preserve lift and drag aerodynamic coefficients is critical to their usefulness as an analysis and design tool. Stratified violin plots of the error distributions are visualized in the center and bottom plots of Fig.~\ref{fig:bwb_reconstruction_dists}. The error is quantified in terms of absolute percent error where sections with absolute values of lift and drag less than $2.5\times10^{-2}$ and $10^{-3}$ respectively are removed to avoid division by small numbers. This filtering process removed  approximately 10\% of the sections for both the lift and drag calculations. The median absolute percent error for lift and drag is 0.3\% and 0.7\% respectively for the POD model, while the median values for LVAE are 4.4\% and 10.6\% respectively. A small number of outliers are present in the MAPE distributions but the majority of errors fall below 1\% and 10\% for the POD and LVAE models respectively. Once again, POD outperforms LVAE on representation transfer to complex 3D aerodynamic shapes. Thus, while LVAE outperforms POD on source, \textit{i.e.}, 2D, dataset compression, the POD model proves more robust to out-of-distribution geometry transfer. Unlike the pressure reconstruction MSE, no strong correlation between lift or drag MAPE and geometry or flow parameters exists. This is attributed to the integrated nature of the calculation where errors in the pressure distribution over the surface may cancel to yield an accurate lift, or drag, predictions. 

\begin{figure}[ht!]
\centering
\includegraphics[width=1.0\textwidth,  trim={3.5cm 0cm 2.5cm 0cm}, clip]{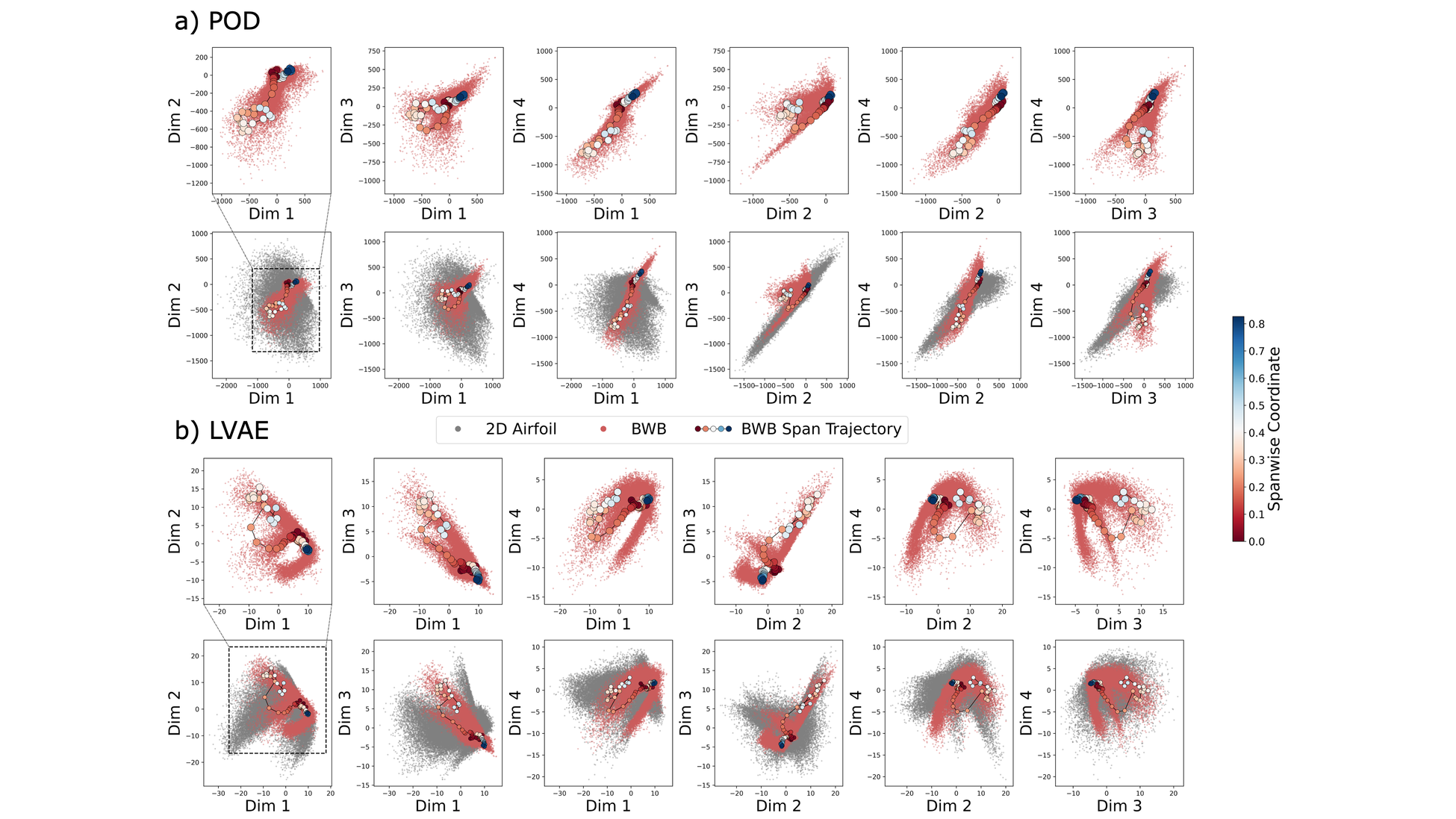}
\caption{Pairwise latent embeddings of 2D airfoil (grey) and BWB (red) data for (a) POD and (b) LVAE. Upper rows magnify the BWB subregion, showing two wing cases with embeddings colored by span coordinate from root (deep red) to tip (navy).}
\label{fig:BWB_latent_embeddings}
\end{figure}

Achieving such low errors across $C_p$, $C_L$, and $C_D$ in this zero-shot transfer setting is a testament to the transfer capability of the POD, and to a lesser extent LVAE, learned representations. We posit that this accuracy can be explained by two factors. First, the 2D pretraining dataset spans a wide range of airfoil geometries, flow conditions, and angles of attack, and contains pressure distributions similar to those produced by relatively complex 3D shapes. Thus a model trained on this comprehensive dataset has been exposed to similar pressure distributions encountered when applied to BWB aircraft. This similarity is demonstrated in Fig.~\ref{fig:BWB_latent_embeddings}, where the low-dimensional BWB representations are proximally embedded near 2D embeddings. However, due to the strong spanwise flow structures present on BWB bodies, the variation of cross-section embeddings with respect to span is notably less smooth than the extruded wing scenario. Second, by learning a low-dimensional manifold, the POD and LVAE models behave as a {\it filter}, in this context projecting 3D BWB data to a 2D flow pattern manifold and filtering out 3D-specific flow features. When applied to 3D data, our low-dimensional representations simply remove the 3D flow features, not captured by the latent dimensions, while preserving the 2D effects. It appears these 2D effects are sufficient to explain the majority of the pressure variations leading to high accuracy reconstructions and preservation of aerodynamic coefficient values.

While the learned representations are clearly challenged by geometric and flow complexities of the blended wing dataset, they achieve notable accuracy on both pressure reconstruction and the integrated aerodynamic coefficients. The accuracy of the POD model in particular suggests the potential for these zero-shot analysis tools to inform design decisions of cutting-edge aircraft in the early design phase where little to no data exists. Furthermore, in the scarce-data regime, limited amounts of 3D data can enhance representation accuracy through fine-tuning that captures 3D-specific flow patterns. We explore this direction in the final results section.

\subsection{\label{sec:results-lvae-finetuning-bwb}3D Fine-tuning}

For real-world design efforts, 3D data can be obtained through high-fidelity numerical fluid simulations or experimental approaches such as wind tunnel testing. Such 3D data possesses rich surface flow information, but collection of the data is hampered by lengthy simulation runs and costly experimentation. Thus, early design phase efforts are restricted by relatively few number of data samples that can be collected. Nevertheless, these samples provide invaluable information regarding the relationship between 3D geometry and aircraft performance. Transfer learning (TL) offers a data efficient approach to incorporate small amounts of 3D data into 2D pretrained models. Thus TL has the potential to achieve higher accuracy representations than could be obtained through pretrained models, as done in Sec. \ref{sec:results-transfer}, or training models solely on sparse 3D data, \textit{i.e.}, training from {\it scratch}. 

To investigate the efficacy of TL within the existing representation learning framework, we employ a dual encoder-decoder strategy. The resulting composite architecture produces two disentangled latent representations, the first pretrained on the large 2D dataset, and the second fine-tuned on a small number (varied from 5 to 500) of 3D surface fields. Specifically, the flow residuals not captured by the 2D representations are used during the fine-tuning process. In concert, the dual encoder-decoder TL model predicts a 2D-informed flow representation and a 3D-informed flow correction which are summed together to produce the final representation. By design, the composite model will produce 3D flow representations which reconstruct BWB pressure distributions to higher accuracy, and with fewer latent dimensions, than a model trained from scratch. The full description of the TL training strategy is presented in App. \S\ref{app:3d_finetuning}.

\begin{figure}[ht!]
\centering
\includegraphics[width=1.0\textwidth]{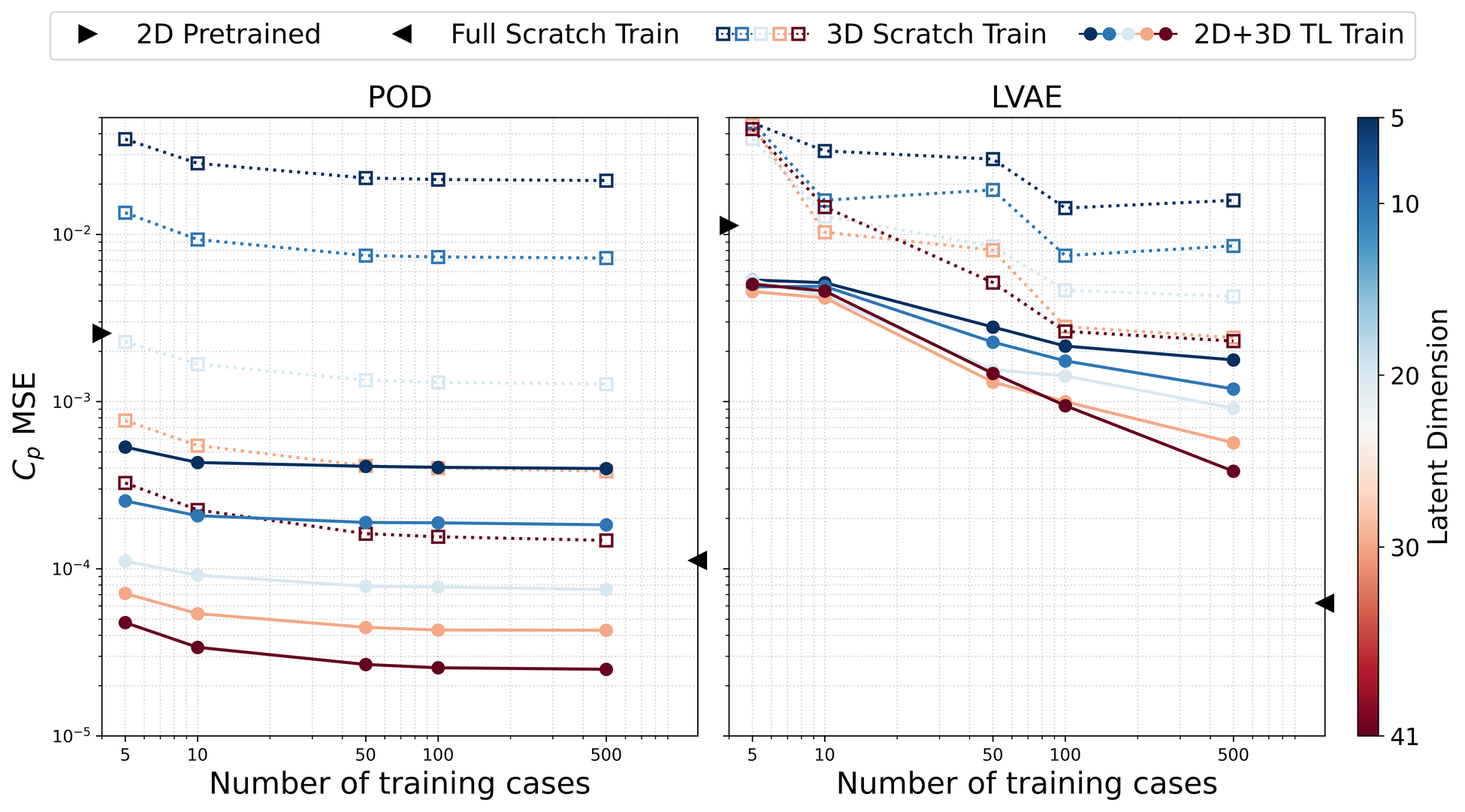}
\caption{Learning curves of $C_p$ test set reconstruction MSE for transfer learning (solid) versus scratch training (dashed) as a function of 3D training set size (5--500 BWB cases). Lines are colored by 3D representation dimensionality (LVAE latent dimensions or POD SVD components).}
\label{fig:tl_scratch_comparison}
\end{figure}

The learning curves are plotted in Fig.~\ref{fig:tl_scratch_comparison} for the POD and LVAE models (left and right subplots, respectively). For each model class, a family of TL and scratch models are trained with varying 3D flow dimensionality, \textit{i.e.}, latent dimensions of an LVAE model or SVD components of a POD model. The TL models are plotted as solid lines while the scratch trained models are dashed. For the POD model, the TL strategy reduces MSE for all dimensions of the 3D flow representation and at all dataset sizes. The TL benefits are greatest in the low-data, compact regime: with just five 3D flow dimensions (38$\times$ compression) and five BWB training cases, TL reduces MSE by 70$\times$ relative to scratch training. As the number of dimensions or number of training cases is increased the error reduction factor decreases but transfer learning {\it always} provides at least a 6$\times$ reduction in error over the scratch training for a fixed latent dimension. 

The LVAE learning curves exhibit similar enhancement of TL models over their scratch-trained counterparts. In fact, TL models capture 3D flow behavior more accurately than the scratch-trained models regardless of the latent dimensionality. Furthermore, TL models display stronger rates of reduction in error with respect to increasing training set size than the scratch models. The relatively higher scratch-trained error and steeper scaling with training set size reflect the overparameterized nature of deep neural architectures such as LVAE, which require sufficient data to constrain their learned mappings and therefore struggle in limited-data regimes. Similar to the POD models, a TL LVAE model restricted to five 3D latent dimensions and five training cases possesses $8.7\times$ lower MSE than the scratch counterpart. 

\begin{figure}[ht!]
\centering
\includegraphics[width=0.9\textwidth]{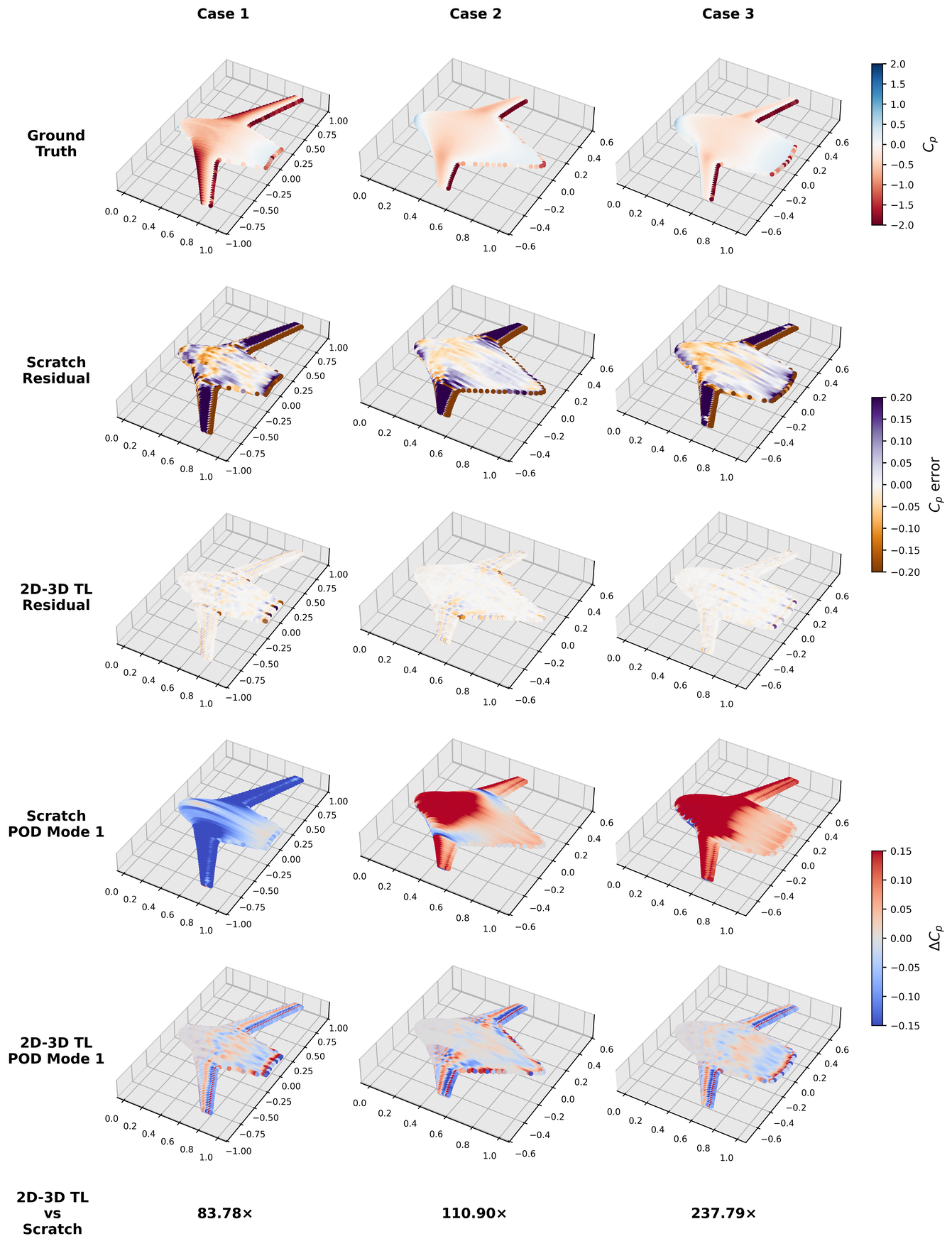}
\caption{Three BWB test cases comparing scratch-trained and TL POD models. Row 1: ground-truth surface pressure. Rows 2--3: pointwise reconstruction residuals for scratch and TL models. Rows 4--5: corresponding 3D flow POD mode 1.}
\label{fig:bwb_flow_representations}
\end{figure}

The TL approach improves model performance not only with respect to scratch training but also with respect to the 2D pretrained model, plotted as a black triangle on the left vertical axis of Fig.~\ref{fig:tl_scratch_comparison}. With only five BWB training cases and five 3D dimensions, TL reduces the MSE by 4.8$\times$ and 2.1$\times$, for POD and LVAE respectively, relative to the 2D pretrained model. For reference, the black triangle on the right vertical axis of Fig.~\ref{fig:tl_scratch_comparison} marks the accuracy achieved by training from scratch on the complete BWB training set, approximately 8,000 cases, using 45 latent dimensions. Consistent with the 2D pretraining scenario, the LVAE model is more accurate than POD when evaluated on a 3D source-domain reconstruction task. Notably, TL POD models with 3D latent dimension greater than 20 outperform full-data scratch training, further supporting the hypothesis that 2D pretraining reduces the complexity of learning 3D flow patterns. The transfer learning strategy thus enables accurate and compact 3D flow representations in the low-data regimes inherent to early phases of aircraft design, where traditional data-driven approaches fail as demonstrated by the poor performance of scratch-trained models (see Fig.~\ref{fig:tl_scratch_comparison} and row 2 of Fig.~\ref{fig:bwb_flow_representations}).

To further examine the benefits conferred by a TL strategy, and interpret the TL-learned 3D flow representations, three distinct BWB cases, drawn from the test set, are examined in each column of Fig.~\ref{fig:bwb_flow_representations}. The TL and scratch approach are applied to POD models trained on just 5 BWB cases and restricted to 5 SVD components, challenging our models in the most compact and data-scarce learning regime. The top row depicts the ground truth surface pressure field while rows two and three show the point-wise residual from a scratch and TL-trained POD model respectively. The scratch POD model produces inaccurate reconstructions, particularly in the wing regions of the BWB body. In contrast, the TL model produces more accurate reconstructions, 84-238$\times$ less MSE per test case, with noticeably lower error in the wing region. Rows 4 and 5 depict POD mode 1 obtained from scratch and TL-training respectively. Note that the scratch trained model POD mode (row 4) varies distinctly between cases, producing a large magnitude decrease of the overall pressure field of Case 1 while increasing the pressure field globally for Cases 2 and 3. Scratch training models are forced to learn 2D and 3D flow effects simultaneously, and the first scratch-trained POD mode appears to correspond global flow distribution features. In contrast, the TL 3D flow POD mode produces pressure corrections primarily along the wing region of the aircraft. Inspection of this mode reveals a clear pattern of alternately decreasing and increasing pressure regions that vary with the chord. For larger versions of these images see App. Fig. ~\ref{fig:tl_pod_mode1_closeup}. In contrast, the flow corrections nearer to the centerline are lower in magnitude as the 2D pretrained model has already captured this {\it 2D-like} flow. Thus, by virtue of the TL approach, TL POD mode 1 can isolate and capture the 3D flow effects not represented by the 2D pretrained model. These 3D flow effects are relatively weaker within the body portion of the geometry and grow in magnitude in the body-wing blending region and pure-wing region. Note that for Case 2, the strongly blended, diamond-like, geometry leads to an onset of 3D flow corrections in the mid-span region. Flow features learned by the TL representations,  \textit{e.g.},the spanwise location of the onset of 3D flow in Case 2, may inform design efforts by connecting this flow behavior to 3D geometric parameters, such as span, sweep, and taper. 

Note that throughout this section, comparisons between the transfer learning and scratch models do not include the 2D pretrained model data requirements and dimensionality. This analysis strategy is motivated by three considerations. First, the computational cost and expertise required to acquire 2D CFD data is typically orders of magnitude less than that required for 3D CFD data, making large, high-quality 2D datasets readily accessible. These datasets provide the foundation for 2D pretrained representations that, as demonstrated in Sec.~\ref{sec:results-transfer}, transfer accurately to geometrically diverse 3D configurations, including extruded wings and BWB aircraft, without any 3D training data. The pretrained models therefore serve as a common starting point for analysis and design across broad classes of aircraft shapes, and the relevant design question becomes how efficiently 3D-specific flow behavior can be learned on top of this foundation. Finally, the disentangled 2D/3D architecture isolates the influence of 3D design choices, such as span, sweep, and taper, on flow behavior, as evidenced by the TL POD modes concentrating corrections in the wing region where 3D effects are strongest (Fig.~\ref{fig:bwb_flow_representations}). Scratch-trained models lack this separation and conflate 2D and 3D flow patterns into a single representation. We therefore present comparisons that highlight the accuracy, data efficiency, and interpretability afforded by fine-tuning 2D pretrained representations versus training from scratch.

\section{\label{sec:conclusion}Conclusion}

This work demonstrates that compact representations of aerodynamic surface fields, learned from readily available 2D airfoil datasets using linear POD and nonlinear LVAE models, transfer accurately to complex 3D aircraft geometries. The 2D pretrained models reconstruct surface pressure fields, lift, and drag coefficients of extruded wing and blended wing body configurations in a zero-shot manner, with POD achieving lift and drag MAPE as low as 0.3\% and 0.7\% on BWB geometries without any 3D training data. When limited 3D data becomes available, a disentangled dual encoder-decoder transfer learning strategy isolates and learns 3D-specific flow patterns, reducing reconstruction error by a factor of 5$\times$ over zero-shot transfer and up to 70$\times$ over scratch training using as few as five 3D cases. The disentangled 3D representations provide actionable design insights, such as identifying the spanwise onset of 3D flow effects and connecting these patterns to geometric parameters including span, sweep, and taper.

A striking finding emerges across all transfer scenarios. While the nonlinear LVAE produces more accurate in-domain reconstructions than POD at matched dimensionality, the linear POD model transfers more accurately to unseen 3D geometry classes in both zero-shot and low-data regimes. This reversal highlights a fundamental distinction between representation compactness and representation transferability that warrants further investigation. 

The extreme data efficiency afforded by transfer learning alleviates the computational bottleneck inherent in design of complex 3D aircraft. Where data-driven design efforts previously required hundreds to thousands of design-specific simulations, the approach presented here learns compact and interpretable representations from readily available 2D datasets that generalize to real-world 3D shapes, enabling modal analysis and design insight in early-phase aircraft design efforts where little to no data exists.

%
%

\section*{Acknowledgements}
The authors gratefully acknowledge project funding provided through the U.S. Army Research Laboratory (ARL), award \# W911NF-23-2-0040.

\section*{Roles}
\textbf{Frank VanGessel:} Conceptualization (equal); Formal analysis (lead); Investigation (lead); Methodology (lead); Data curation (lead); Software (lead); Validation (lead); Visualization (lead); Writing - original draft (lead); Writing - review \& editing (lead). \textbf{Cashen Diniz:} Data curation (supporting); Writing - review \& editing (supporting). \textbf{Mark Fuge:} Conceptualization (equal); Methodology (supporting); Formal analysis (supporting)

\section*{Data Availability}
The extruded wing data that support the findings of this study were made openly available by the original authors at https://github.com/cashend/OptiWing3D. The blended wing data that support the findings of this study were made openly available by the original authors in Harvard Dataverse at https://doi.org/10.7910/DVN/VJT9EP.

\clearpage

\appendix

\section{\label{app:data_processing}Data Processing}

All airfoil, wing section, and blended-wing-body datasets are standardized to a uniform discretization prior to training using two interpolation routines. Geometry coordinates are resampled by fitting a B-spline of degree $k=3$ to the raw coordinate sequence and resampling $N=192$ points according to curvature-weighted arc length. Specifically, a cumulative curvature integral is computed along a high-resolution parametric curve (1000 evaluation points), augmented by a uniform shift constant $D=20$ to prevent excessive point clustering in low-curvature regions. New parameter values are then sampled uniformly over the augmented cumulative curvature, and final coordinates are evaluated from the fitted spline at those parameter locations. This scheme concentrates points near regions of high curvature (e.g., the leading edge).

Surface pressure field values are remapped from the original mesh onto the resampled geometry using a radial basis function (RBF) interpolator with a thin-plate spline kernel. The interpolation is consistent across all dataset classes — airfoils, wings, and blended wing bodies. The radial basis function interpolation is implemented using the Scikit-learn library~\cite{scikit-learn} using the default interpolation values.

\section{\label{app:lvae_arch}LVAE Model Architecture and Training Strategy}

\subsection{Encoder Architecture}

The encoder is a convolutional neural network one-dimensional encoder that maps a discretized pressure field
$\mathbf{p} \in \mathbb{R}^{1 \times 192}$ to a latent vector
$\mathbf{z} \in \mathbb{R}^{300}$.
It consists of five strided 1-D convolutional layers followed by a two-layer
multilayer perceptron (MLP):

\begin{enumerate}
    \item \textbf{Convolutional front-end.}
    Five \texttt{Conv1d} layers with kernel size 4, stride 2, and padding 1
    progressively expand the channel dimension while halving the spatial
    resolution at each step:
    $(1, 192) \to (64, 96) \to (128, 48) \to (256, 24) \to (512, 12) \to (1024, 6)$.
    Each layer is followed by a \textit{Leaky ReLU} activation with negative
    slope $\alpha = 0.2$.
    The output is flattened to a vector of size $1024 \times 6 = 6144$.
    \item \textbf{MLP head.}
    Two fully-connected hidden layers of widths $[1024, 512]$, each followed by
    Leaky ReLU ($\alpha = 0.2$), project the flattened features to the
    $d_0 = 300$-dimensional initial latent space.
    No activation is applied after the final linear layer.
\end{enumerate}

\subsection{Decoder Architecture}

The decoder is a spectrally normalized one-dimensional deconvolutional neural network that maps $\mathbf{z} \in \mathbb{R}^{300}$
back to a pressure field $\hat{\mathbf{p}} \in \mathbb{R}^{1 \times 192}$.
All weight matrices are subject to \textit{spectral normalization}~\cite{miyato2018spectral}
to enforce Lipschitz normalization.

\begin{enumerate}
    \item \textbf{MLP stem.}
    A two-hidden-layer spectrally-normalized MLP with widths $[512, 1024]$
    maps $\mathbf{z}$ to a vector of size $1024 \times 6 = 6144$, which is
    reshaped into a feature map of shape $(1024, 6)$.
    Each hidden layer uses Leaky ReLU ($\alpha = 0.2$).
    \item \textbf{Transposed-convolutional back-end.}
    Five spectrally-normalized \texttt{ConvTranspose1d} layers, each with
    kernel size 4, stride 2, and padding 1, double the spatial resolution
    while reducing the channel depth:
    $(1024, 6) \to (512, 12) \to (256, 24) \to (128, 48) \to (64, 96) \to (1, 192)$.
    The first four layers are followed by Leaky ReLU ($\alpha = 0.2$);
    the final layer has no activation, producing the reconstructed pressure
    field directly.
\end{enumerate}

\subsection{Loss Function}

The model is trained with a composite loss consisting of a reconstruction term
and a volume regularization term:
\begin{equation}
    \mathcal{L} = w_{\mathrm{rec}}\,\mathcal{L}_{\mathrm{rec}}
                + w_{\mathrm{vol}}\,\mathcal{L}_{\mathrm{vol}},
    \label{eq:lvae_loss}
\end{equation}
where
\begin{equation}
    \mathcal{L}_{\mathrm{rec}} = \frac{1}{N}\|\mathbf{p} - \hat{\mathbf{p}}\|^2,
    \label{eq:rec_loss}
\end{equation}
is the mean-squared reconstruction error and
\begin{equation}
    \mathcal{L}_{\mathrm{vol}}
        = \exp\!\left(\frac{1}{d}\sum_{i=1}^{d}\log \sigma_i \right)
        = \left(\prod_{i=1}^{d} \sigma_i\right)^{\!1/d}
    \label{eq:vol_loss}
\end{equation}
is the geometric mean of the per-dimension standard deviations
$\sigma_i = \mathrm{std}_{b}\bigl[z_i\bigr]$ of the active latent dimensions
over a mini-batch, with $d$ denoting the current number of active (unpruned)
dimensions.
The volume loss penalizes the \emph{hypervolume} of the latent distribution,
driving the encoder to concentrate information into as few dimensions as
possible~\cite{chen2025volumeanalysis}.

\subsection{Loss Weight Schedule}

The volume loss weight is annealed via a polynomial schedule to allow the
reconstruction to stabilize before regularization takes effect:
\begin{equation}
    w_{\mathrm{vol}}(e) =
    \begin{cases}
        0, & e < 200,\\[4pt]
        w \left(\dfrac{e - 200}{200}\right)^{2}, & 200 \le e < 400,\\[6pt]
        w, & e \ge 400,
    \end{cases}
    \label{eq:schedule}
\end{equation}
with $w = 10^{-6}$.
The reconstruction weight is held constant at $w_{\mathrm{rec}} = 1$ throughout.
The model is trained for 3000 epochs with a mini-batch size of 64 using the
Adam optimizer~\cite{kingma2014adam} with learning rate $10^{-4}$.

\subsection{Dynamic Pruning}

After a warm-up period of 500 epochs, latent dimensions are pruned at the end
of each epoch.
During training the encoder maintains exponential moving averages of the
per-dimension standard deviation $\tilde{\sigma}_i$ and mean $\tilde{\mu}_i$
with momentum $\beta = 0.9$.
A \emph{plummet ratio} is computed by sorting dimensions by $\tilde{\sigma}$
in descending order and identifying the largest gap in $\log\tilde{\sigma}$;
all dimensions whose ratio relative to the plummet dimension falls below a
threshold $\tau = 0.02$ are pruned:
\begin{equation}
    \rho_i = \frac{\tilde{\sigma}_i}{\tilde{\sigma}_{\mathrm{plummet}}}, \quad
    \text{prune dimension } i \;\text{if}\; \rho_i < \tau.
    \label{eq:prune}
\end{equation}
Pruned dimensions are frozen at their current moving-mean values $\tilde{\mu}_i$
and excluded from the volume loss computation.
The volume loss is renormalized by the fraction of active dimensions
$m = d / d_0$ to prevent the objective from collapsing as dimensions are removed.

\subsection{Validation Checkpointing}

Model checkpoints are saved using a validation-based checkpoint manager that
monitors the reconstruction loss on a held-out validation set.
A checkpoint is saved when either (i) the validation reconstruction loss
strictly decreases, or (ii) the active latent dimension $d$ decreases relative
to the previously saved checkpoint and the validation loss does not exceed the
current best by more than 20\%. The best checkpoint (lowest validation reconstruction loss) is used for all
reported results.

\section{\label{app:lvae_training_curves}LVAE Training Curves}

\begin{figure}[htbp]
    \centering
    \begin{subfigure}{\textwidth}
        \centering
        \includegraphics[width=1.0\textwidth]{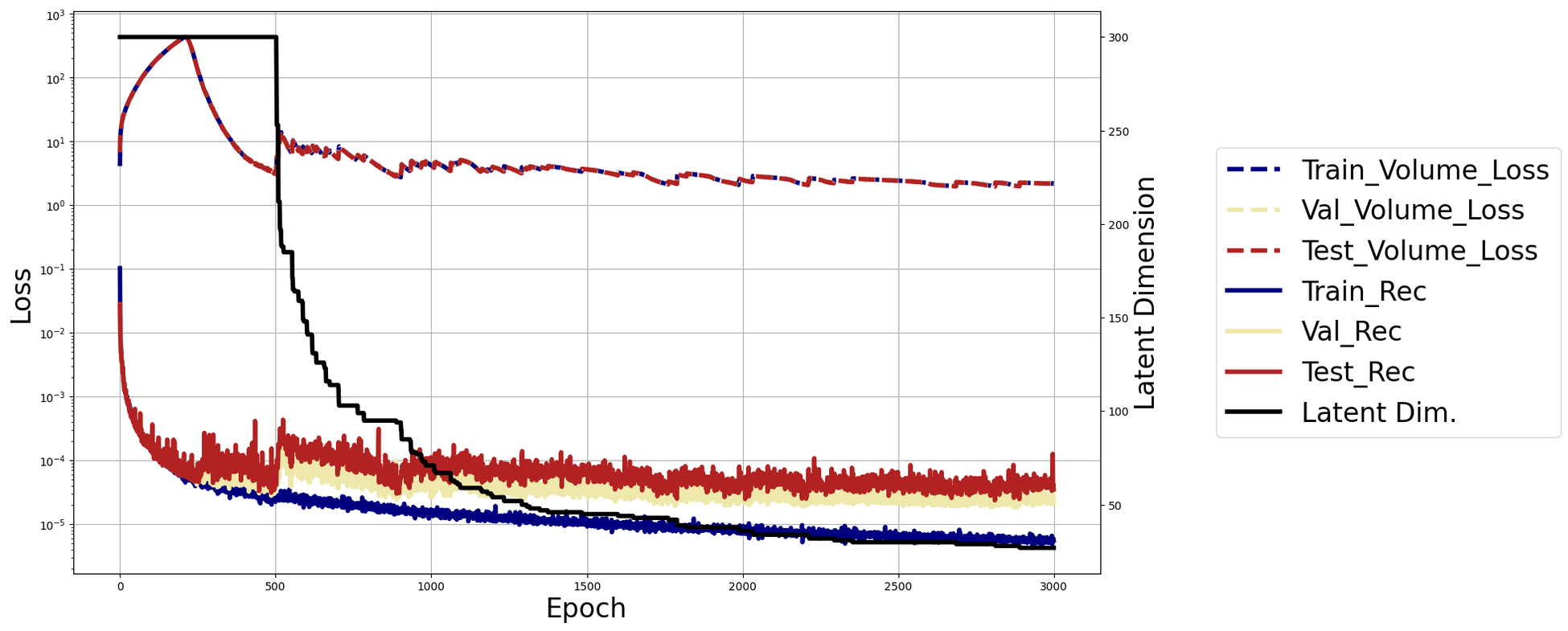}
        \caption{2D Dataset 1.}
        \label{fig:loss_ds1}
    \end{subfigure}

    \begin{subfigure}{\textwidth}
        \centering
        \includegraphics[width=1.0\textwidth]{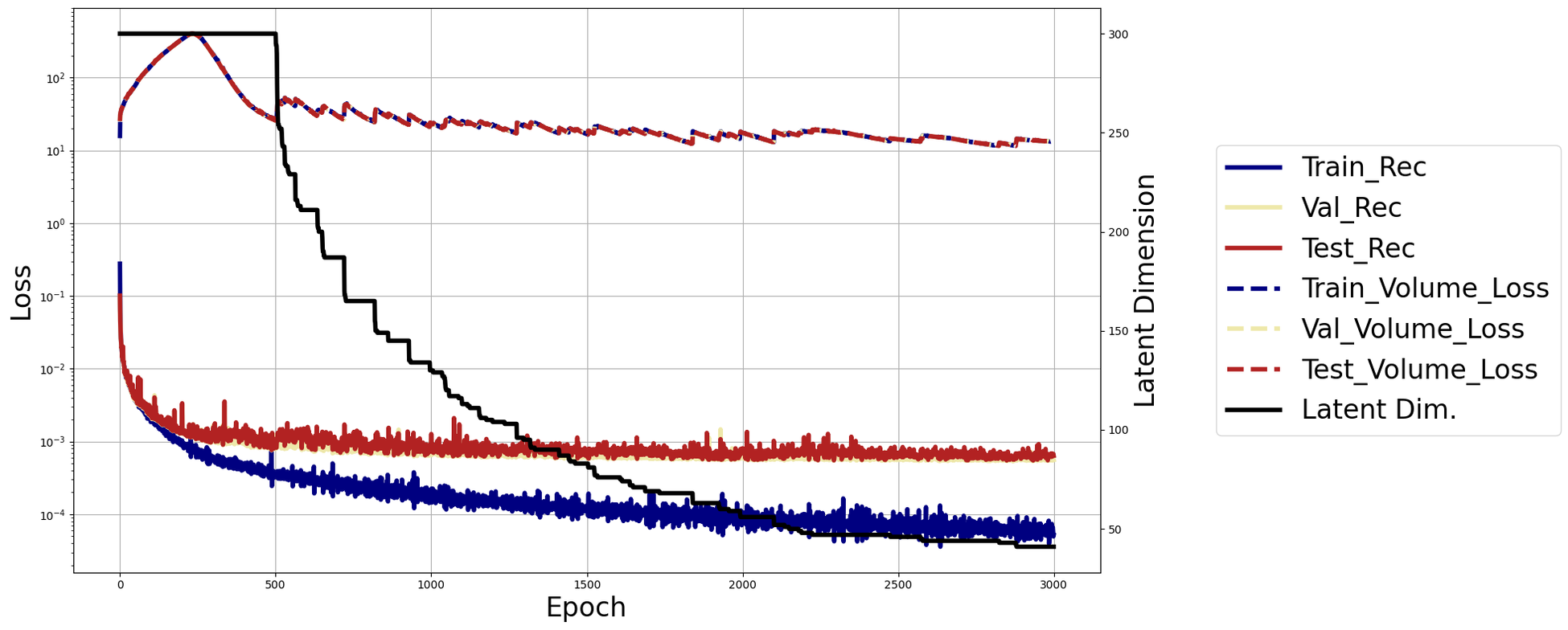}
        \caption{2D Dataset 2.}
        \label{fig:loss_ds2}
    \end{subfigure}
    \caption{Training loss curves for LVAE model.}
\label{fig:lvae_loss}
\end{figure}

The training curves for loss and dimensionality versus training epoch for 2D Dataset 1 \& 2 are given in Fig.~\ref{fig:lvae_loss}. Notice that the training curves show the initial warm-up of the volume penalty over the first 500 epochs, after which dimensional pruning begins. Subsequently, the model proceeds to balance accurate reconstruction loss with a reduction in latent space volume and dimensionality. The final model achieves both a compact latent representation and excellent reconstruction accuracy. The training curves for Dataset 2 exhibit a similar pattern.

\section{\label{app:latent_interpolation}Latent Interpolation}

The latent interpolation plots for POD and LVAE models trained on Dataset 1 and 2 are presented in Fig.~\ref{fig:latent_interp_all}.

\begin{figure}[htbp]
    \centering
    \begin{subfigure}{\textwidth}
        \centering
        \includegraphics[width=\linewidth]{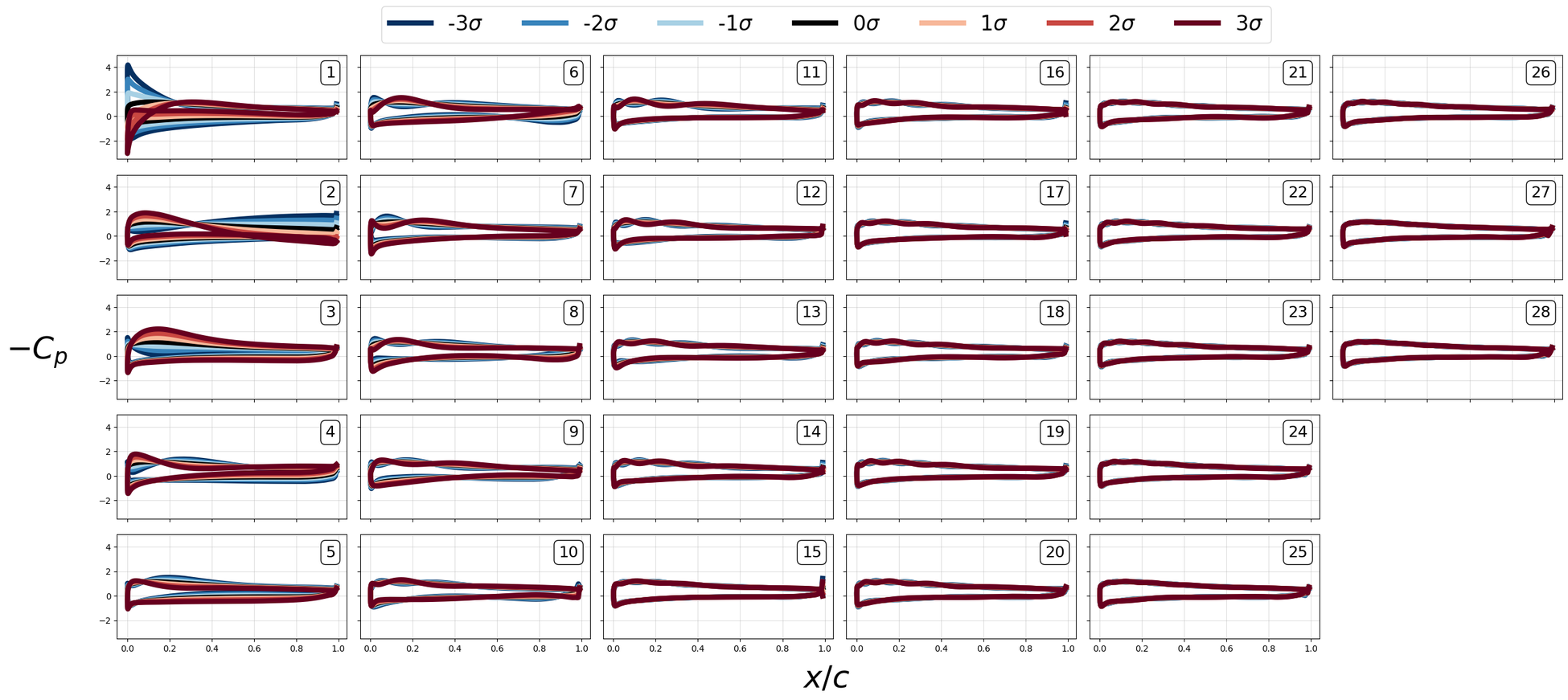}
        \caption{POD latent interpolation for 2D Dataset 1.}
        \label{fig:latent_interp_pod_DS1}
    \end{subfigure}

    \begin{subfigure}{\textwidth}
        \centering
        \includegraphics[width=\linewidth]{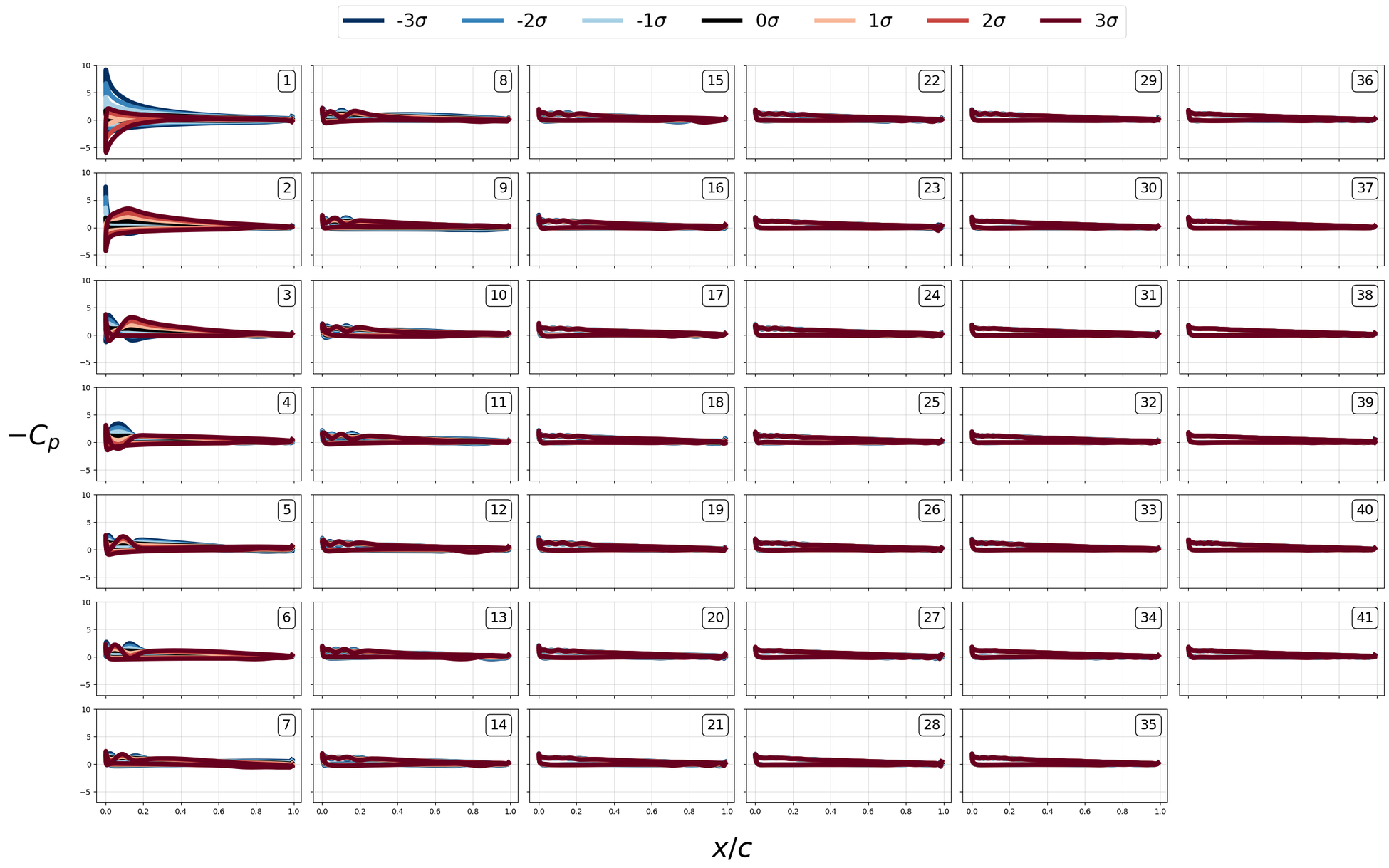}
        \caption{POD latent interpolation for 2D Dataset 2.}
        \label{fig:latent_interp_pod_DS2}
    \end{subfigure}
    \caption{Reconstructed pressure fields corresponding to latent interpolation about the mean latent embedding for all latent dimensions of both 2D Datasets. The reconstructed pressure is plotted with respect to average $x$-coordinate. The pressure field reconstructed from the latent mean is plotted as a black curve. The subplots are numbered according to the importance of the latent dimension to the surface pressure field reconstruction.}
    \label{fig:latent_interp_all}
\end{figure}

\begin{figure}[htbp]
    \ContinuedFloat
    \centering
    \begin{subfigure}{\textwidth}
        \centering
        \includegraphics[width=\linewidth]{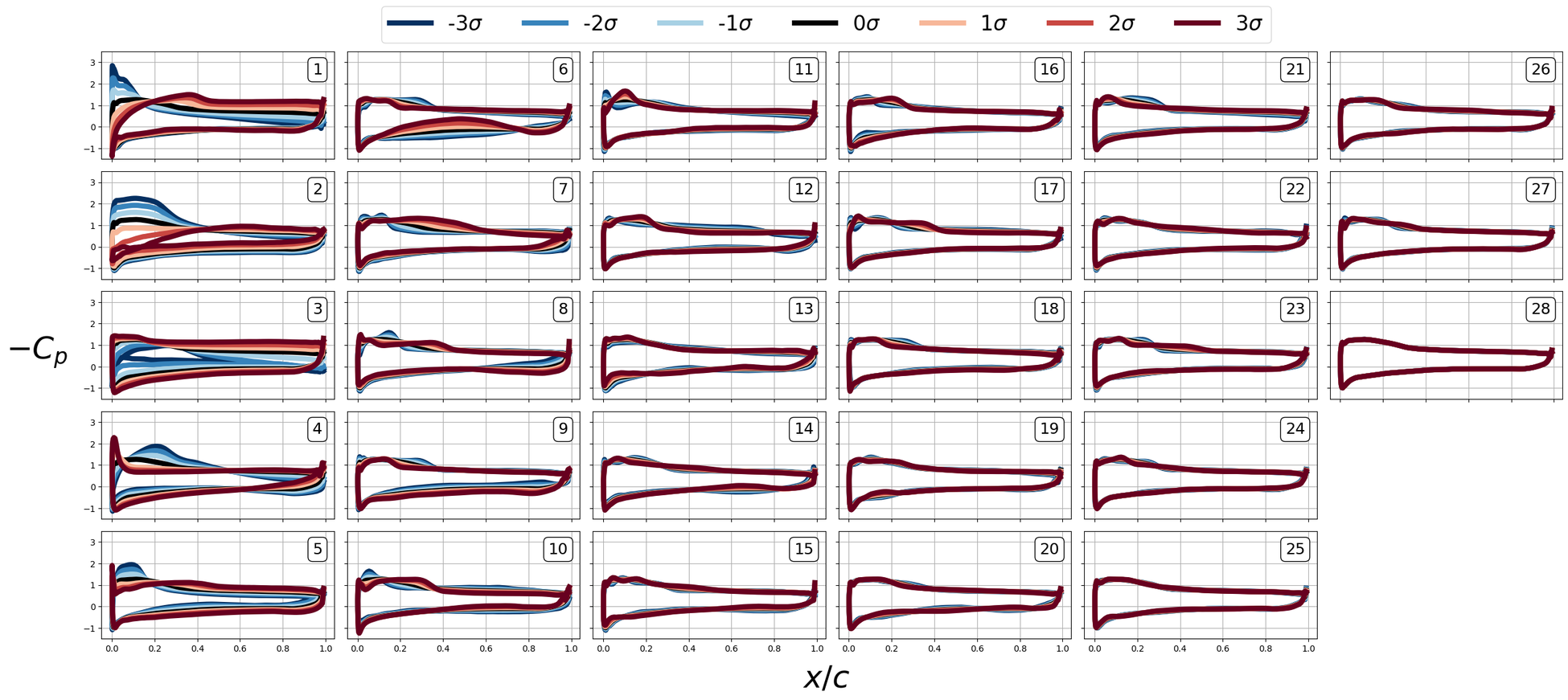}
        \caption{LVAE latent interpolation for 2D Dataset 1.}
        \label{fig:latent_interp_lvae_DS1}
    \end{subfigure}

    \begin{subfigure}{\textwidth}
        \centering
        \includegraphics[width=\linewidth]{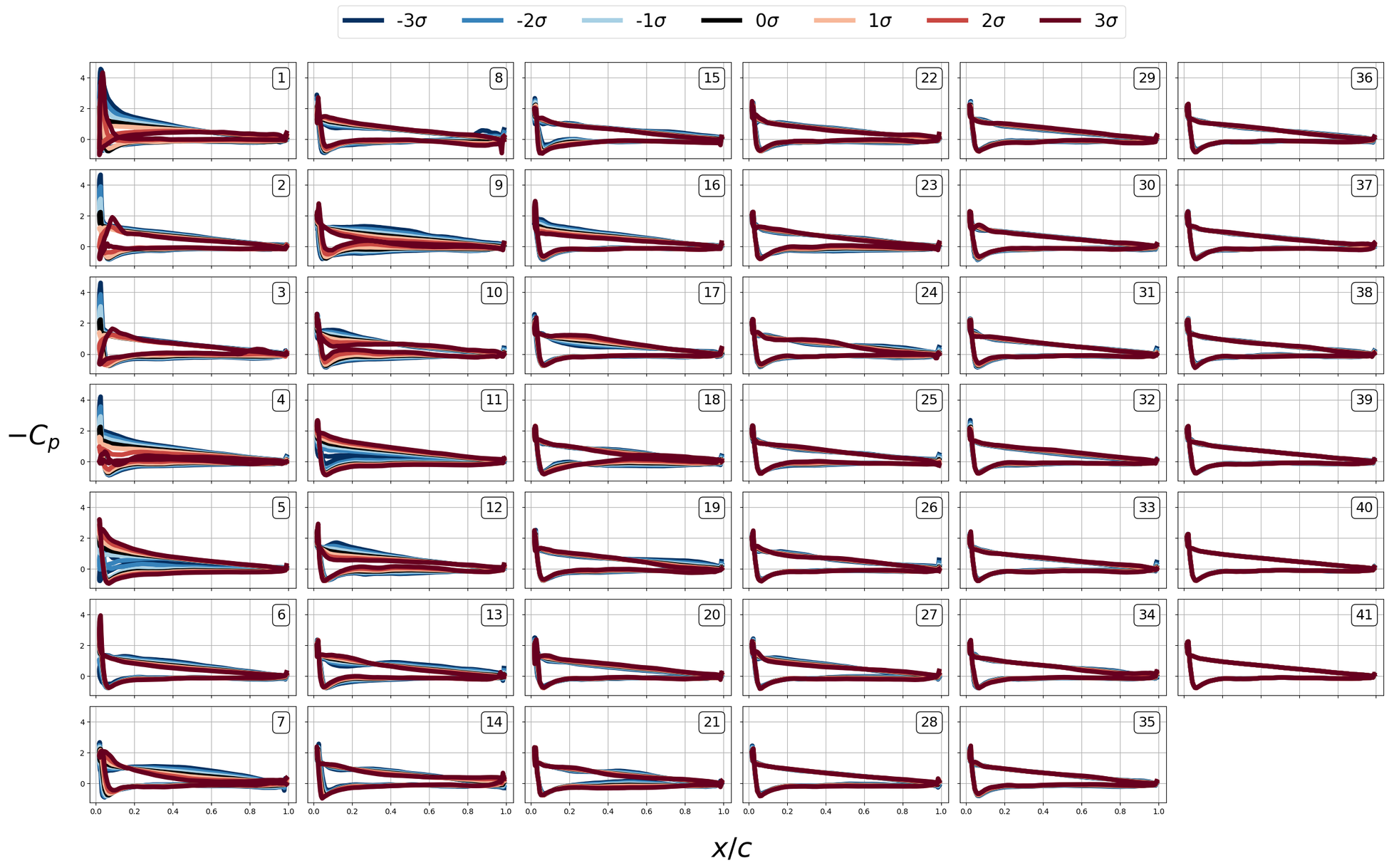}
        \caption{LVAE latent interpolation for 2D Dataset 2.}
        \label{fig:latent_interp_lvae_DS2}
    \end{subfigure}
    \caption*{Figure~\thefigure. (Continued)}
\end{figure}

\section{Composite Transfer Learning Architecture and Training Procedure}
\label{app:3d_finetuning}

\subsection{Overview}

The composite transfer learning (TL) approach combines a fixed pretrained 2D~LVAE
(described in Appendix ~\ref{app:lvae_arch}) with a second LVAE trained exclusively on
the prediction residuals of the 3D blended-wing-body (BWB) dataset.
Given a 3D pressure field $\mathbf{f} \in \mathbb{R}^{1 \times 192}$ at a single
spanwise section, the ensemble prediction is
\begin{equation}
    \hat{\mathbf{f}}_{\mathrm{ens}}
        = \hat{\mathbf{f}}_{\mathrm{pre}}
        + \hat{\mathbf{f}}_{\mathrm{res}}\!\left(
              \widetilde{\mathbf{r}}
          \right),
    \label{eq:ensemble}
\end{equation}
where $\hat{\mathbf{f}}_{\mathrm{pre}}$ is the pretrained model's
reconstruction, $\mathbf{r} = \mathbf{f} - \hat{\mathbf{f}}_{\mathrm{pre}}$ is
the pointwise residual, and $\hat{\mathbf{f}}_{\mathrm{res}}$ is the residual
model's reconstruction of the standardized residual
$\widetilde{\mathbf{r}} = (\mathbf{r} - \mu_r)/\sigma_r$.
The constants $\mu_r$ and $\sigma_r$ are the global mean and standard deviation
computed from the training-split residuals and are fixed thereafter.
The residual model's output is inverse-scaled before addition:
$\hat{\mathbf{f}}_{\mathrm{res}} \leftarrow \hat{\mathbf{f}}_{\mathrm{res}}\,\sigma_r + \mu_r$.
The pretrained model's weights are frozen throughout residual training.

\subsection{Residual Model Architecture}

The residual LVAE shares the identical encoder--decoder architecture as the
pretrained model (Conv1DEncoder and TrueSNDCGenerator1D, see
Appendix ~\ref{app:lvae_arch}), with the same initial latent dimension $d_0 = 300$.
The key distinction is the use of the \emph{minimum-dimension} constraint. Namely,
the \texttt{min\_prune\_dim} parameter sets a lower bound $d_{\min}$ on the
number of active latent dimensions that the dynamic pruning mechanism may retain.
If a pruning action would reduce the number of active dimensions below $d_{\min}$,
further pruning is suppressed and training continues with exactly $d_{\min}$
dimensions active. This allows systematic evaluation of reconstruction accuracy as a function of a
\emph{fixed} latent dimension, aiding comparison among LVAE residual models of varied dimension and POD models of the same fixed dimension. Experiments are conducted at $d_{\min} \in \{5, 10, 20, 30, 41\}$.

\subsection{Residual Model Training}

The residual LVAE is trained with the same composite loss as the pretrained
model (Eq.~\ref{eq:lvae_loss}), but with a compressed warm-up:
\begin{equation}
    w_{\mathrm{vol}}(e) =
    \begin{cases}
        0, & e < 50,\\[4pt]
        w \left(\dfrac{e - 50}{50}\right)^{2}, & 50 \le e < 100,\\[6pt]
        w, & e \ge 100,
    \end{cases}
    \label{eq:res_schedule}
\end{equation}
where $w$ is the volume loss weight determined by hyperparameter search
(Section~\ref{app:tl_hparam}).
Training runs for 5000~epochs with mini-batch size~64, Adam optimizer, and
learning rate $10^{-4}$.
The dynamic pruning momentum is $\beta = 0.9$ and pruning begins at epoch~100.
Checkpoints are selected by lowest validation reconstruction loss, with a
tolerance of $10\times$ on validation loss increase when the active dimension
decreases, reflecting the noisier convergence expected with small 3D datasets.

\subsection{Hyperparameter Selection Strategy}
\label{app:tl_hparam}

Two hyperparameters govern the residual LVAE's training dynamics: the volume
loss weight $w$ and the pruning ratio threshold $r_t$.
Because the optimal values depend on the size $N$ of the 3D training set,
hyperparameters are tuned separately for each value of $N \in \{5, 10, 50, 100, 500\}$
cases.
For a given $N$, a grid search over $(w,\,r_t)$ is performed with the minimum
dimension fixed at $d_{\min} = 5$. The combination achieving the lowest test-set MSE of the composite ensemble
(Eq.~\ref{eq:ensemble}) is selected. This optimal $(w^*,\,r_t^*)$ pair is then held fixed while training residual models at the remaining fixed dimensions $d_{\min} \in \{10, 20, 30, 41\}$ for
the same training set size $N$.

\subsection{POD Composite Baseline}

An analogous composite approach is applied using Proper Orthogonal Decomposition
(POD) as the linear counterpart to the LVAE. The 2D pretrained POD model is used as a 2D basis.
The 3D residuals $\mathbf{r} = \mathbf{f} - \hat{\mathbf{f}}_{\mathrm{pre}}$
are projected onto a second POD basis fitted from the $N$ 3D training cases.
The composite prediction is formed identically to Eq.~\ref{eq:ensemble}.
To ensure a fair comparison, the number of residual POD modes is set to match
the fixed latent dimension $d_{\min}$ of the corresponding LVAE residual model,
\textit{i.e.}, \ $n_{\mathrm{components}} \in \{5, 10, 20, 30, 41\}$.
Because POD has no trainable hyperparameters beyond the component count, no
grid search is required; the component count is determined entirely by the LVAE
grid search described above.

\subsection{2D-3D TL POD Mode 1 Figures}

\begin{figure}[ht!]
\centering
\includegraphics[width=\textwidth]{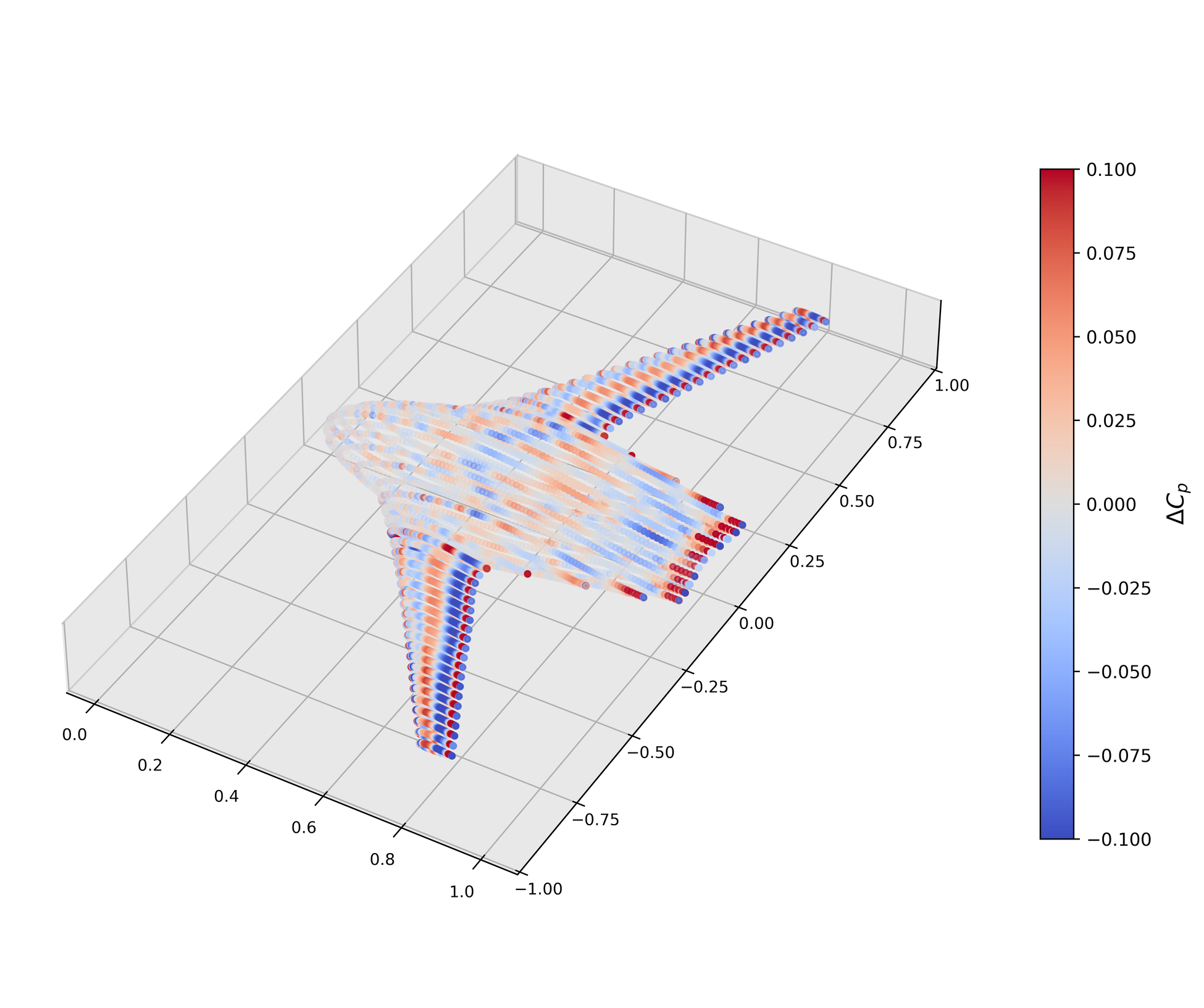}
\caption{Transfer learned POD Mode 1 for three BWB test cases. Case 1.}
\label{fig:tl_pod_mode1_closeup}
\end{figure}

\begin{figure}[ht!]
\ContinuedFloat
\centering
\includegraphics[width=\textwidth]{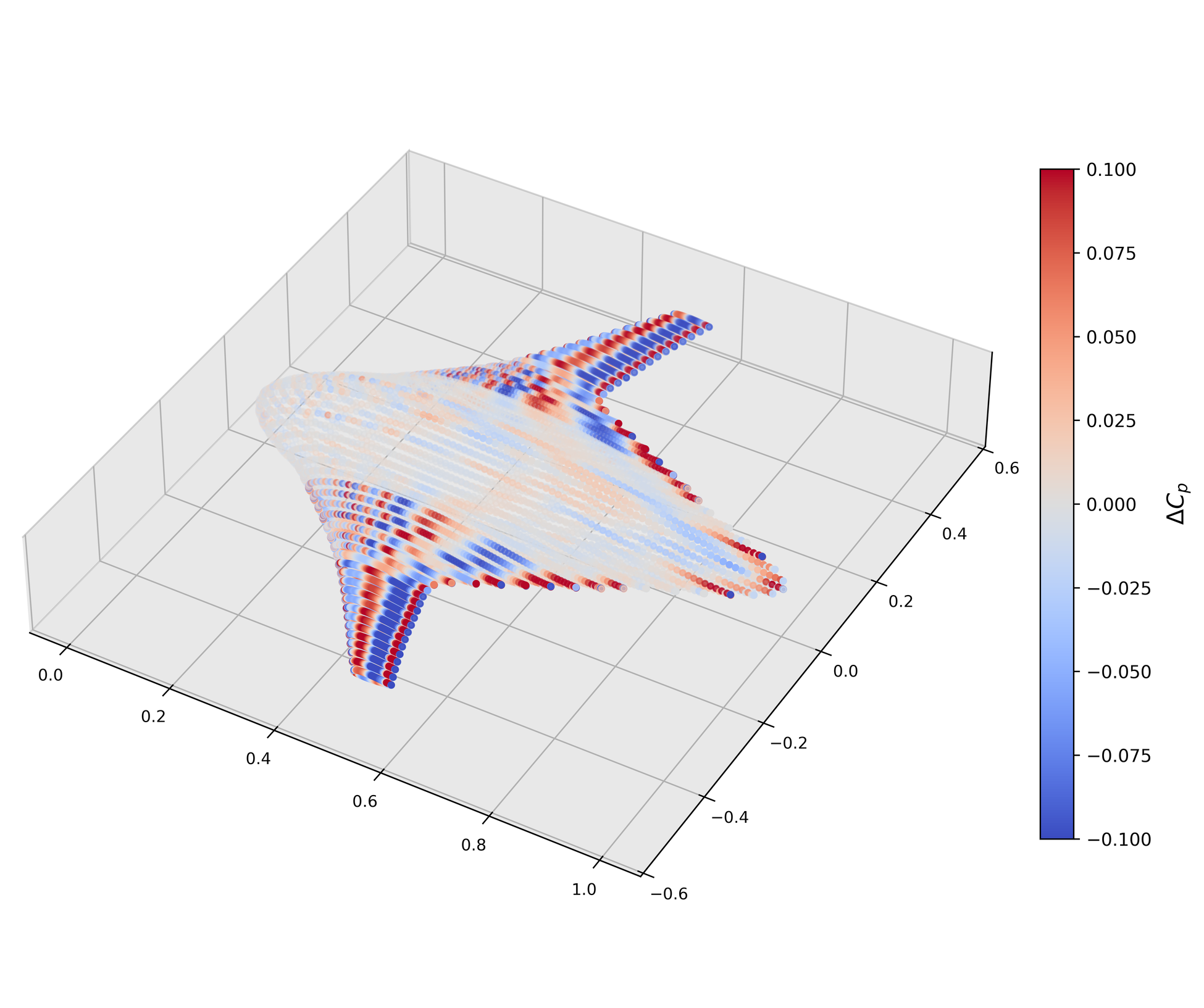}
\caption*{Figure~\thefigure. (Continued) Case 2.}
\end{figure}

\begin{figure}[ht!]
\ContinuedFloat
\centering
\includegraphics[width=\textwidth]{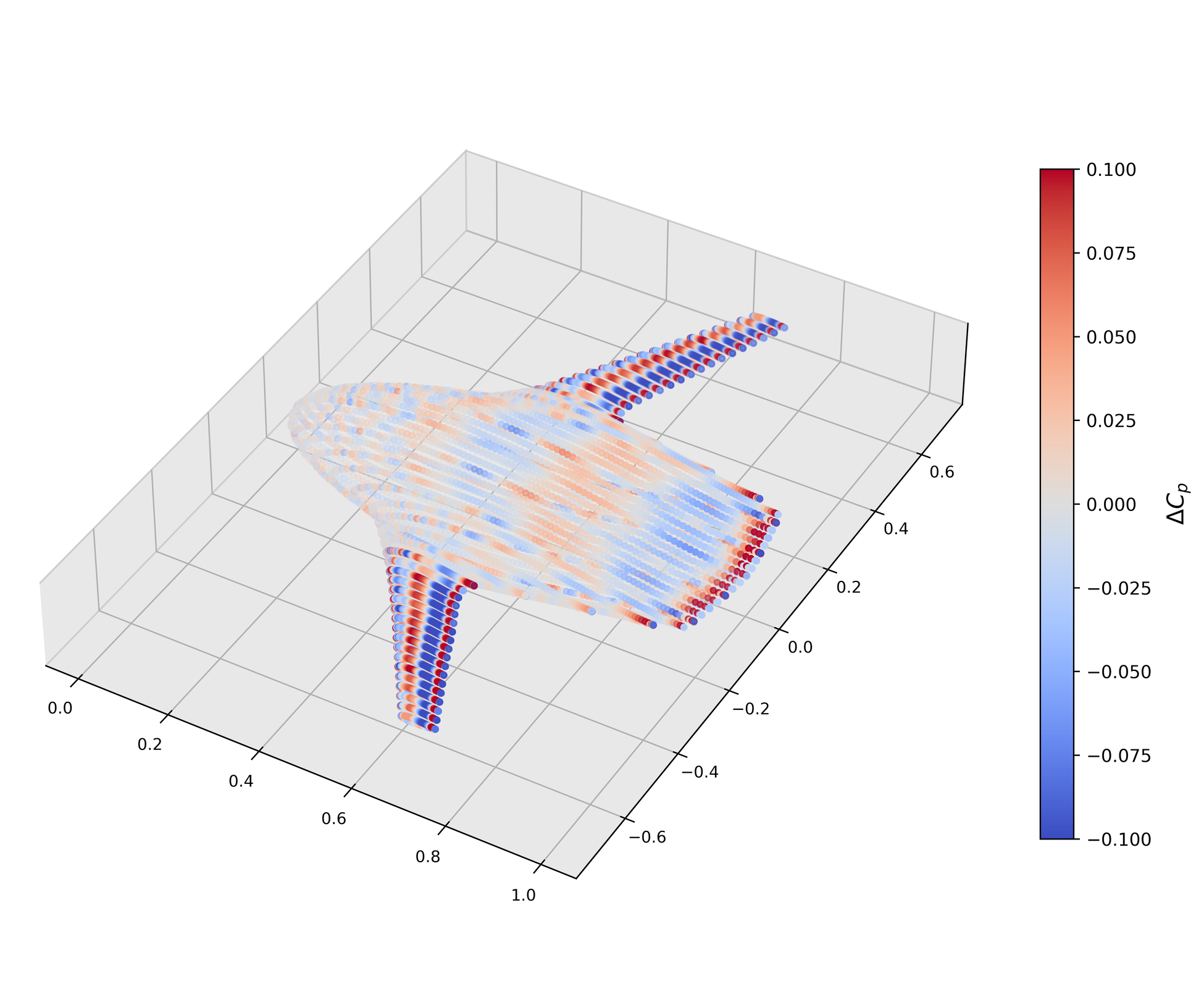}
\caption*{Figure~\thefigure. (Continued) Case 3.}
\end{figure}

\clearpage

\bibliographystyle{iopart-num}
\bibliography{aipsamp}

\end{document}